\documentclass[11pt,a4paper]{article}
\usepackage{epsfig}
\usepackage{a4wide}
\usepackage{latexsym}
\usepackage{graphicx}
\usepackage{amsmath}
\usepackage{slashed}
\usepackage{amsfonts}
\usepackage{amssymb}
\usepackage{mathrsfs}
\usepackage{appendix}
\usepackage[usenames,dvips]{color}
\usepackage[colorlinks=true,urlcolor=PineGreen,anchorcolor=PineGreen,citecolor=PineGreen,filecolor=PineGreen,linkcolor=PineGreen,menucolor=PineGreen,pagecolor=PineGreen,linktocpage=true,pdfproducer=medialab]{hyperref}
\usepackage{layout}

\allowdisplaybreaks[4]
\numberwithin{equation}{section}

\begin{document}

\begin{titlepage}

\begin{center}
{\Large {\textbf{{Geometry of Special Galileon}}}}\\[1.5 cm]%
\textbf{ Ji\v{r}\'{\i} Novotn\'y}\footnote{
for email use: novotny at ipnp.troja.mff.cuni.cz},  \\[1 cm]
\textit{Institute of Particle and Nuclear Physics, Faculty of
Mathematics
and Physics,}\\[0pt]
\textit{Charles University, V Hole\v{s}ovi\v{c}k\'ach 2, CZ-180 00
Prague 8,
Czech Republic} \\[0.5 cm]
\end{center}

\begin{abstract}
Theory known as Special Galileon has recently attracted considerable interest due to its peculiar properties.  It has been shown  that it represents an extremal member of the set of effective field theories with enhanced soft limit. This property makes its tree-level $S-$matrix fully on-shell reconstructible and representable by means of the Cachazo-He-Yuan representation. The enhanced soft limit is a consequence of new hidden symmetry of the Special  Galileon action, however, until now, the origin of this peculiar  symmetry has remained unclear.  In this paper we interpret this symmetry as a special transformation of the coset space  $GAL\left( D,1\right) /SO(1,D-1)$ and show, that there exists a three-parametric family of invariant Galileon actions. The latter family is closed under duality which appears as a natural generalization of the above mentioned symmetry. We also present a geometric construction of the Special Galileon action using $D-$dimensional brane propagating in $2D-$dimensional flat pseudo-riemannian space. Within such framework, the Special Galileon symmetry emerges as an $U(1,D-1)$ symmetry of the target space, which can be treated as a $D-$dimensional  K\"{a}hler manifold. Such a treatment allows for classification of the higher order invariant Lagrangians needed as counterterms on the quantum level. We also briefly comment on relation between such higher order Lagrangians and the Lagrangians invariant with respect to the polynomial shift symmetry.
\end{abstract}

\end{titlepage}

\tableofcontents

\bigskip \line(1,0) {100}

\setcounter{footnote}{0}
%%%%%%%%%%%%%%%%%%%%%%%%%%%%%%%%%%%%%%%%%%%%%%%%%%%%%%%%%%%%%%%%%%%%

\section{Introduction and summary of the results}

Galileons are known to be very interesting derivatively coupled real scalar
field theories with a rich spectrum of applications. Originally the simplest
cubic Galileon emerged as an effective field theory describing the only
interacting scalar mode in a decoupling limit of the Dvali-Gabadadze-Porrati
model \cite{Dvali:2000hr,Deffayet:2001pu}. It also naturally appeared in
analogous decoupling limit of theory with massive graviton \cite%
{deRham:2010kj} where it described a zero-helicity mode. It has been soon
realized, that generalizations of the cubic Galileon exist, namely $D$
dimensions allow for $D+1$ independent Galileon Lagrangian terms. Such a
general Galileon theory has been then proposed as a long distance
modification of the General relativity \cite{Nicolis:2008in}. An appealing
feature of Galileon in this context is a presence of Vainshtein screening
\cite{Vainshtein:1972sx} as well as a stability of its basic Lagrangian with
respect to the quantum corrections \cite%
{Luty:2003vm,Hinterbichler:2010xn,deRham:2012ew}. Other generalizations as a
vector or $p-$form Galileons and its covariant form on a curved background
has been discussed e.g. in \cite{Deffayet:2009mn,Deffayet:2010zh}.

There exist also close relation between the Galileon theories and theories
describing fluctuations of $D$ dimensional brane in $D+1$ dimensional
Minkowski space-time. Within this framework the flat space Galileon emerges
as a certain non-relativistic limit of special versions of such theories
\cite{deRham:2010eu,Hinterbichler:2010xn}. This approach allowed for further
generalizations changing the Minkowski space for Anti-de Sitter space or for
a general curved background and recovering in this way the conformal
Galileon and covariant Galileon respectively.

Recently the Galileon theories attracted further attention due to the
special distinctive properties of their $S-$matrix. It has been found \cite%
{Cheung:2014dqa,Cheung:2015ota,Cheung:2016drk}, that the general Galileon is
a unique theory (up to the values of the couplings and within a certain
class of single scalar theories with specific power-counting) the amplitudes
of which have non-trivial one-particle soft limit. More precisely, let us
write the interaction Lagrangian of a single scalar effective theory
schematically as%
\begin{equation}
\mathcal{L}_{int}=\sum_{d,n}\lambda _{d,n}\partial ^{d}\phi ^{n}
\end{equation}%
and let us restrict ourselves to the theories with vertices for which%
\begin{equation}
\rho \equiv \frac{d-2}{n-2}=2.
\end{equation}%
Then the general Galileon is the only theory whose scattering amplitudes $%
A_{m}\left( p,\ldots \right) $ vanish as the second power of momentum,
\begin{equation}
A_{m}\left( p,\ldots \right) \overset{p\rightarrow 0}{=}O\left( p^{2}\right)
\end{equation}%
when one of the particles in the $in$ or $out$ state becomes soft. Moreover,
within the class of the general Galileon theories there exists a
distinguished one (dubed \emph{Special Galileon}), which exhibits even more
enhanced soft behavior - the corresponding amplitudes vanish as the third
power of the soft momentum \cite{Cheung:2014dqa,Cachazo:2014xea}. The latter
feature makes the Special Galileon very peculiar: its $S-$matrix can be
fully reconstructed from the lowest nontrivial on-shell amplitude i.e. form
the four-point one\footnote{%
Also the $S-$matrix of the general Galileon in $D$ dimensions is on-shell
reconstructible \cite{Cheung:2015ota}, however, as an input, it is necessary
to know all the on shell amplitudes up to the $D+1$ point one.}\cite%
{Cheung:2015ota,Cheung:2016drk}. Also, the $S-$matrix of the Special
Galileon has the Cachazo-He-Yuan representation \cite{Cachazo:2014xea}
(along with other exceptional scalar field theories with enhanced soft
limit, namely the $U\left( N\right) $ non-linear sigma model and the
Dirac-Born-Infeld theory (DBI)). As discussed in \cite{Cheung:2016drk}, all
these theories occupy the border line $\rho =\sigma -1$ which delimits the
allowed region for theories with the non-trivial soft limit%
\begin{equation}
A_{n}\left( p\right) \overset{p\rightarrow 0}{=}O\left( p^{\sigma }\right)
\end{equation}%
in the $\left( \sigma ,\rho \right) $ plane and so their soft behavior is
extremal. The Special Galileon is in fact \textquotedblleft doubly
extremal\textquotedblleft\ because is sits in a corner of the allowed region
and its soft exponent $\sigma =3$ is the highest possible one \cite%
{Cheung:2016drk}.

The soft behavior of the scattering amplitudes is closely related to the
non-linear symmetries of the underlying theory \cite{Cheung:2016drk}. For
all the above mentioned exceptional scalar field theories there exist a
symmetry of the action of the type%
\begin{equation}
\delta _{\theta }\phi \left( x\right) =\theta _{\alpha _{1}\ldots \alpha
_{n}}\left[ x^{\alpha _{1}}\ldots x^{\alpha _{n}}+\Delta ^{\alpha _{1}\ldots
\alpha _{n}}\left( x\right) \right] ,
\end{equation}%
where $\theta _{\alpha _{1}\ldots \alpha _{n}}$ are infinitesimal parameters
and where $\Delta ^{\alpha _{1}\ldots \alpha _{n}}\left( x\right) $ is a
linear combination of local composite operators. The very presence of such a
symmetry guaranties, under some regularity assumptions, that the soft
exponent satisfies $\sigma \geq n+1$. For instance the $O(p^{2})$ soft limit
of the general Galileon is a consequence of the linear shift symmetry%
\begin{equation}
\delta _{\theta }^{Gal}\phi \left( x\right) =\theta _{\alpha }x^{\alpha },
\end{equation}%
and the same $\sigma =2$ soft behavior of the Dirac-Born-Infeld theory is a
consequence of the symmetry%
\begin{equation}
\delta _{\theta }^{DBI}\phi \left( x\right) =\theta _{\alpha }\left[
x^{\alpha }-F^{-D}\phi \left( x\right) \partial ^{\alpha }\phi \left(
x\right) \right]  \label{DBI_symmetry}
\end{equation}%
Let us remind, that within the latter theory the field $\phi \left( x\right)
$ describes a position of the brane in the extra dimension. The symmetry (%
\ref{DBI_symmetry}) of the DBI action has then a nice geometrical
interpretation as a non-linearly realized Lorentz transformation of the
target $D+1$ dimensional space in which the fluctuating brane propagates.
This geometrical picture is very useful because it simplifies considerably
the constructions of higher order invariant Lagrangians (i.e. those with
more than one derivative per field). The latter are necessary as couterterms
when the higher loop corrections are taken into account. The very
identification of the DBI symmetry as a group of invariance of the target
space metric converts the task of counterterm construction in an almost
routine enumeration of the reparameterization invariants built from the
induced metric on the brane and its extrinsic curvature. In this sense the DBI theory and its symmetry
is fully understood.

This is not the case of the Special Galileon, although a hidden symmetry
responsible for its $O\left( p^{3}\right) $ soft behavior has been found
shortly after discovery of this peculiar theory \cite{Hinterbichler:2015pqa}%
. This symmetry can be written in the form%
\begin{equation}
\delta _{\theta }^{sGal}\phi \left( x\right) =\theta ^{\mu \nu }\left(
\Lambda x_{\mu }x_{\nu }-\partial _{\mu }\phi \left( x\right) \partial _{\nu
}\phi \left( x\right) \right)  \label{sGal_symmetry}
\end{equation}%
where $\theta ^{\mu \nu }$ is a symmetric traceless tensor and $\Lambda $ is
a parameter related to the coupling constant of the Special Galileon.
However, the origin and properties of this symmetry have been still unclear and
its form does not allow for construction of higher order countertems in an
easy and straightforward way. Even the direct proof that the transformation (%
\ref{sGal_symmetry}) is really a symmetry of the Special Galileon action is
rather complicated, because the contributions of different parts of the
Lagrangian with different number of fields has to cancel each other in a
subtle way.

The aim of this article is to elucidate this issue. We will first show, that
the hidden symmetry (\ref{sGal_symmetry}) has its origin in a certain finite
reparameterization of the coset space $GAL\left( D,1\right) /SO(1,D-1)$. The
latter corresponds to the spontaneous symmetry breaking of the Galileon
symmetry
\begin{equation}
\delta _{c,d}\phi \left( x\right) =c+d_{\mu }x^{\mu },
\label{general_Gal_symmetry}
\end{equation}%
according to the pattern\footnote{%
Here, as above, $D$ is the space-time dimension.} $GAL\left( D,1\right)
\rightarrow ISO\left( 1,D-1\right) $. As it has been shown in \cite%
{Goon:2012dy}, the general Galileon Lagrangian can be recognized as a linear
combination of Schwinger terms related to closed $GAL\left( D,1\right) $%
-invariant $D+1$ forms $\omega _{D+1}^{\left( n\right) }$ on the above
mentioned coset. This coset space can be parameterized by $2D+1$ coordinates
$\phi $ , $L_{\mu }$ and $x^{\mu }$, which are related to the generators of
the general Galileon symmetry (\ref{general_Gal_symmetry}) and space-time
translations. The general Galileon Lagrangian in then obtained by means of
integrating the forms $\omega _{D+1}^{\left( n\right) }$over $D+1$
dimensional ball $B_{D+1}$ whose boundary is the compactified space-time and
imposing then the inverse Higgs constraint $L_{\mu }=\partial _{\mu }\phi $.

We will show that the coset space reparametrization $\left( \phi
,~L,~x\right) \rightarrow \left( \phi ,~L_{\theta },~x_{\theta }\right) $,
where%
\begin{eqnarray}
x_{\theta }^{\mu } &=&\left[ \cosh \left( 2a\theta \right) \right] _{~\nu
}^{\mu }x^{\nu }+\frac{1}{a}\left[ \sinh \left( 2a\theta \right) \right]
_{~\nu }^{\mu }L^{\nu }  \notag \\
L_{\theta }^{\mu } &=&a\left[ \sinh \left( 2a\theta \right) \right] _{~\nu
}^{\mu }x^{\nu }+\left[ \cosh \left( 2a\theta \right) \right] _{~\nu }^{\mu
}L^{\nu }  \notag \\
\phi _{\theta } &=&\phi -\frac{1}{2}L^{\mu }x_{\mu }+\frac{1}{2}L_{\theta
}^{\mu }\cdot (x_{\theta})_{\mu }  \label{coset_sGal_transformation}
\end{eqnarray}%
is responsible for the hidden Galileon symmetry (\ref{sGal_symmetry}) with $%
\Lambda =a^{2}$. Here the matrices $\cosh \left( 2a\theta \right) $ and $%
\sinh \left( 2a\theta \right) $ are corresponding functions\footnote{%
In the powers of $\theta $ the indices are contracted with the flat metric
tensor.} of the symmetric tensor $\theta ^{\mu \nu }$ which is a free
parameter of the transformation. The case $\Lambda <0$ can be obtained by
means of analytic continuation $a\rightarrow \mathrm{i}\alpha $. We will
also show, that for fixed $a$ there is a two-parametric family\footnote{%
In the analytically continued case $a=\mathrm{i}\alpha $ and $\alpha $ fixed
there is only one-parameter family due to the requirement of reality of the
action.} of actions, $S\left( a,c_{+},c_{-}\right) $ which are invariant
with respect to (\ref{coset_sGal_transformation}) for traceless $\theta
^{\mu \nu }$. Moreover, in the case when $\theta ^{\mu \nu }$ is not
traceless, the transformation (\ref{coset_sGal_transformation}) becomes a
duality of the family $S\left( a,c_{+},c_{-}\right) $, namely such actions
transform according to%
\begin{equation}
S\left( a,c_{+},c_{-}\right) \rightarrow S\left( a,c_{+}+a\theta _{\mu
}^{\mu },c_{-}-a\theta _{\mu }^{\mu }\right) .
\end{equation}%
The theory known in the literature as a Special Galileon then corresponds 
(up to an overall normalization) to the case $c_{+}=c_{-}$. We have
therefore not only hidden Galileon symmetry, but also a \emph{hidden
Galileon duality }of the whole family of Special Galileons.

Our second result shows, that the above symmetry/duality has a nice
geometrical origin. Namely, we will prove that the Galileon field can be
interpreted as a scalar degree of freedom which describes position of
special $D-$dimensional brane in $2D$ dimensional flat space with
pseudo-riemannian metric whose signature is either $(D,D)$ or $(2,2D-2)$.
The coordinates on such a $2D$ space can be identified with the above
mentioned coset space coordinates $x^{\mu }$ and $L_{\mu }$ and the metric
reads%
\begin{equation}
\mathrm{d}s^{2}=\eta _{\mu \nu }\left( \mathrm{d}X^{\mu }\mathrm{d}X^{\nu }-%
\frac{1}{\Lambda }\mathrm{d}L^{\mu }\mathrm{d}L^{\nu }\right) .
\end{equation}%
The inverse Higgs constraint $L_{\mu }=\partial _{\mu }\phi $ is implemented
demanding the brane to be isotropic with respect to the properly chosen
two-form, namely%
\begin{equation}
\omega =\eta _{\mu \nu }\mathrm{d}X^{\mu }\wedge \mathrm{d}L^{\nu }.
\end{equation}%
The $\omega -$preserving subgroup of isometries of the bulk then acts on the
brane configurations, which are described solely by one scalar field $\phi $%
, and result in nonliner transformations which are exactly the general
Galileon symmetry (\ref{general_Gal_symmetry}) (the translations in the $%
L_{\mu }$ directions) and the hidden Galileon duality (\ref%
{coset_sGal_transformation}) (rotations which mix $x^{\mu }$ and $L_{\mu }$%
). As a consequence and similarly to the construction of the DBI action, the
invariant Galileon action can be constructed as brane action which consists
of reparameterization invariants build form geometric objects like the
induced metric on the brane%
\begin{equation}
g_{\mu \nu }=\eta _{\mu \nu }-\frac{1}{\Lambda }\partial _{\mu }\partial
_{\alpha }\phi \partial ^{\alpha }\partial _{\nu }\phi ,
\label{induced_sGal_metric}
\end{equation}%
or the brane extrinsic curvature. The higher order Lagrangian terms
invariant with respect to (\ref{coset_sGal_transformation}), which are
needed as counterterms for the special Galileon, can be then easily
classified. We make such a classification up to the terms of the schematic
structure $\partial ^{2n+4}\phi ^{n}$ and $n$ arbitrary.

Our third results concerns the relation between the higher order special Galileon
Lagrangians and the Lagrangians invariant with respect to the polynomial
shift symmetries. The latter have been studied and classified in \cite%
{Griffin:2014bta} and \cite{Hinterbichler:2014cwa}. We will show, that in $%
D>2n$ dimensions the quadratic shift invariants of the general form $%
\partial ^{6n+4}\phi ^{2n+2}$ (i.e. $\left( P,N,\Delta \right) =\left(
2,2n+2,3n+2\right) $ when using the notation\footnote{%
Here $P$ is the order of the polynomial shift, $N$ is a number of legs and $%
2\Delta $ is nuber of derivatives.} of \cite{Griffin:2014bta}) can be easily
obtained from our construction. Namely, provided we restrict ourselves to
the Lovelock action \cite{Lovelock:1971yv} built form $n-$th power of the
Riemann tensor corresponding to the metric (\ref{induced_sGal_metric}), we
can obtain the invariant $\left( 2,2n+2,3n+2\right) $ just summing up the
all the terms with $N=2n+2$ legs.

The paper is organized as follows. In the Section 2 we briefly review a
coset construction of the general Galileon action and fix our notation. In
Section 3 we shortly discuss the relation between transformations of the
coset and general Galileon dualities. In Section 4 we introduce a special
coset space transformation which generates a duality of certain
three-parametric family of Galileon actions and the infinitesimal form of
which is identical with the hidden symmetry of Special Galileon for special
choice of the parameters. Here we also discuss possible generalizations of
the above mentioned duality/symmetry. Sections 5 and 6 are devoted to the
geometrical construction of the Special Galileon using the probe brane in $%
2D-$dimensional space. In Sections 7 and 8 we present a construction of the
higher order Lagrangians and their classification up to the terms $%
\partial^{2n+4}\phi^n$. Here we also comment on the relation to the actions
invariant with respect to the quadratic shift. In Section 9 we briefly
discuss the second branch of the Special Galileon symmetry within the probe
brane context. Conclusions are summarized in Section 10. Some technical
details are postponed to appendices. In Appendix A we prove the invariance
of the Inverse Higgs Constraint with respect to the special Galileon
duality. The same is done for generalized special Galileon duality in
Appendix B. In Appendix C we give an explicit form of some higher order
Lagrangian terms. In Appendix D we discuss the properties of the family of
invariant Lagrangians under other duality transformations.

\section{Coset construction of the Galileon action}

In this section we give a brief overview\footnote{%
For more detailed treatment we refer to the original paper \cite{Goon:2012dy}%
, as an example of further application see also \cite{Kampf:2014rka}.} of
the geometry of the coset space behind the Galileon Lagrangian and its
connection to the more traditional treatment. It appears that such a
geometrical language is particularly useful and elegant for investigation of
additional symmetries and dualities of the theory while the traditional
approach makes these aspects less transparent. In what follows we will also
fix our conventions and notation.

It is well known that the Galielon field $\phi \left( x\right) $ can be
understood as a Goldstone boson of the spontaneously broken Galileon%
\footnote{%
The Galileon symmetry is the simplest nontrivial example of the general
polynomial shift symmetry discussed in \cite{Hinterbichler:2014cwa} and \cite%
{Griffin:2014bta}.} symmetry\footnote{%
Here and in what follows the dot means a contraction of the adjacent Lorentz
indices, i.e.%
\begin{equation*}
L\cdot x=L^{\mu }x_{\mu }=L_{\mu }x^{\mu }=L^{\mu }\eta _{\mu \nu }x^{\nu
}=L\cdot \eta \cdot x.
\end{equation*}%
In the same spirit, the double (multiple) dot between symmetric tensors
means double (multiple) contraction, e.g.%
\begin{equation*}
\partial \partial \phi \colon \eta =\partial _{\mu }\partial _{\nu }\phi
\eta ^{\mu \nu },~~\partial \partial \partial \phi \vdots \partial \partial
\partial \phi ~=\partial _{\mu }\partial _{\nu }\partial _{\rho }\phi
\partial ^{\mu }\partial ^{\nu }\partial ^{\rho }\phi
\end{equation*}%
e.t.c..}
\begin{equation}
\delta _{a,b}\phi \left( x\right) =a+b\cdot x,  \label{linear_shift}
\end{equation}%
according to the symmetry breaking pattern%
\begin{equation}
GAL\left( D,1\right) \rightarrow ISO\left( 1,D-1\right) .
\end{equation}%
Here $GAL\left( D,1\right) $ is the Galileon group in $D$ space-time
dimensions with generators \ $P_{a}$, $J_{ab}=-J_{ba}$ (space-time
translations, rotations and boosts) and $A$, $B_{a}$ (constant shift and
non-uniform linear shift of the field respectively). These generators
satisfy the algebra
\begin{eqnarray}
\left[ P_{a},P_{b}\right] &=&[B_{a},B_{b}]=[P_{a},A]=[B_{a},A]=\left[
J_{ab},A\right] =0  \notag \\
\lbrack P_{a},B_{b}] &=&\mathrm{i}\eta _{ab}A  \notag \\
\lbrack J_{ab},P_{c}] &=&\mathrm{i}\left( \eta _{bc}P_{a}-\eta
_{ac}P_{b}\right)  \notag \\
\lbrack J_{ab},B_{c}] &=&\mathrm{i}\left( \eta _{bc}B_{a}-\eta
_{ac}B_{b}\right)  \notag \\
\lbrack J_{ab},J_{cd}] &=&\mathrm{i}\left( \eta _{bc}J_{ad}+\eta
_{ad}J_{bc}-\eta _{ac}J_{bd}-\eta _{bd}J_{ac}\right) .  \label{GAL}
\end{eqnarray}%
In terms of the generators $A$ and $B_{a}$ we can write the infinitesimal
Galileon transformation (\ref{linear_shift}) as%
\begin{equation}
\delta _{a,b}=-\mathrm{i}aA-\mathrm{i}b^{a}B_{a}.
\end{equation}%
The symmetry breaking order parameter is the vacuum expectation value
\begin{equation}
\left\langle 0|\delta _{a,b}\phi |0\right\rangle =a+b\cdot x.
\label{order_parameter}
\end{equation}

As it has been recognized in \cite{Goon:2012dy}, the basic (i.e. the lowest
order\footnote{%
For the systematic power counting scheme for the Galileon see \cite%
{Kampf:2014rka}$.$ According to this counting, the operators are classified
using the index $\delta =d-2(n-1)$ where $d$ is number of derivatives and $n$
is number of the external lines. The lowest order Galileon Lagrangian
corresponds to the linear combination of the operators with $\delta =0$.})
Galileon Lagrangian in $D$ dimensional space-time represents a linear
combination of $D+1$ generalized Wess-Zumino-Witten terms \cite%
{Wess:1971yu,Witten:1983tw,D'Hoker:1994ti,D'Hoker:1995it,D'Hoker:1995it}.
These can be constructed using the standard coset construction of Callan,
Coleman, Wess and Zumino \cite{Coleman:1969sm,Callan:1969sn} which has been
adapted for non-uniform symmetries\footnote{%
That means those symmetries which do not comute with the spacetime
translations.} by Volkov \cite{Volkov:1973vd} and Ivanov and Ogievetsky \cite%
{Ivanov:1975zq}.

In the Galileon case the coset space is\footnote{%
Note that the only generators which are realized linearly on the field space
are the rotations and boosts $J_{ab}$.}
\begin{equation}
GAL\left( D,1\right) /SO(1,D-1)=\{gSO(1,D-1),g\in GAL\left( D,1\right) \}
\label{non-linear realization}
\end{equation}%
and can be parametrized by means of the coordinates $x^{\mu }$, $L^{\mu }$
and $\phi $ (the latter two are candidates for the Goldstone fields)
corresponding to the following choice of the representative $U$ of each left
coset
\begin{equation}
U=\mathrm{e}^{\mathrm{i}P\cdot x}\mathrm{e}^{\mathrm{i}A\phi +\mathrm{i}%
B\cdot L}\in \left\{ \mathrm{e}^{\mathrm{i}P\cdot x}\mathrm{e}^{\mathrm{i}%
A\phi +\mathrm{i}B\cdot L}SO(1,D-1)\right\} .
\end{equation}%
This choice induces a non-linear realization of the transformation $g\in
GAL\left( D,1\right) $ on the coset according to the prescription%
\begin{equation}
\left\{ \mathrm{e}^{\mathrm{i}P\cdot x^{^{\prime }}}\mathrm{e}^{\mathrm{i}%
A\phi ^{^{\prime }}+\mathrm{i}B\cdot L^{^{\prime }}}SO(1,D-1)\right\}
=\left\{ g\mathrm{e}^{\mathrm{i}P\cdot x}\mathrm{e}^{\mathrm{i}A\phi +%
\mathrm{i}B\cdot L}SO(1,D-1)\right\} .
\end{equation}%
The basis of the covariant building blocks (dubbed $\omega _{A}$, $\omega
_{P}^{a}$ and $\omega _{B}^{a}$ in what follows) which are used for the
construction of the $GAL\left( D,1\right) $ invariant action can be then
read off from the Maurer-Cartan form%
\begin{equation}
\frac{1}{\mathrm{i}}U^{-1}\mathrm{d}U=\omega _{P}^{c}P_{c}+\omega
_{A}A+\omega _{B}^{d}B_{d}.
\end{equation}%
Explicitly we get%
\begin{eqnarray}
\omega _{A} &=&\mathrm{d}\phi -L^{a}\eta _{ab}\cdot \mathrm{d}x^{b}  \notag
\\
\omega _{P}^{a} &=&\mathrm{d}x^{a}  \notag \\
\omega _{B}^{a} &=&\mathrm{d}L^{b}.  \label{forms}
\end{eqnarray}%
Lorentz invariants constructed from the components of the forms $\omega _{A}$%
, $\omega _{P}^{a}$ and $\omega _{B}^{a}$ and the ordinary\footnote{%
The covariant derivative $\nabla _{\mu }$ used in the general Volkov and
Ogievetsky construction is in this case simply the ordinary derivative $%
\partial _{\mu }$.} derivative $\partial _{\mu }$ can be shown to be
automatically invariant with respect to the non-linear realization (\ref%
{non-linear realization}) of $GAL\left( D,1\right) $ on the coset.

Note that the localized form of the Galileon symmetry with space-time
dependent parameters $a(x)$ and $b(x)$%
\begin{equation}
\delta _{a(x),b(x)}\phi =a(x)+b(x)\cdot x\equiv \widehat{a}(x)=\delta _{%
\widehat{a}(x)}\phi
\end{equation}%
looks like the localization of the constant shift with parameter $\widehat{a}%
(x)$. Therefore it is not possible to distinguish between the localization
of the constant shift and the linear shift, and as a cosequence, there is
only one physical Goldstone boson $\phi \left( x\right) $ corresponding to
the local fluctuation of the order parameter (\ref{order_parameter}) \cite%
{Low:2001bw,McArthur:2010zm,Watanabe:2011ec,Brauner:2014aha}. The unphysical
Goldstones $L^{\mu }\left( x\right) $ can be eliminated imposing the Inverse
Higgs Constraint (IHC) \cite{Ivanov:1975zq} which in this case reads%
\begin{equation}
\omega _{A}=0.  \label{IHC}
\end{equation}%
This implies%
\begin{equation}
\omega _{B}^{a}=\mathrm{d}L^{\mu }\left( x\right) =\partial ^{\mu }\partial
_{\nu }\phi \left( x\right) \mathrm{d}x^{\nu }.
\end{equation}%
The building blocks for the invariant Lagrangian are therefore the second
and higher derivatives of the Galileon field. The most general invariant
Lagrangian is then
\begin{equation}
\mathcal{L}_{\mathrm{inv}}=\mathcal{L}_{\mathrm{inv}}(\partial _{\mu
}\partial _{\nu }\phi ,\partial _{\lambda }\partial _{\mu }\partial _{\nu
}\phi ,\ldots ).
\end{equation}%
It is manifestly invariant with respect to the Galileon symmetry because the
number of derivatives acting on each field is sufficient to compensate the
linear shift (\ref{linear_shift}). However such a Lagrangian is in fact
higher order in the power counting with respect to the basic Galileon
Lagrangian. The latter has smaller number of derivatives per field,
schematically%
\begin{equation}
\mathcal{L}_{\mathrm{basic}}=\sum_{n=1}^{D+1}\partial ^{2n-2}\phi ^{n}.
\end{equation}%
This is possible because the basic Lagrangian is invariant only up to a
total derivative and therefore the apparently deficient number of
derivatives which do not compensate the linear shift of all the fields is in
fact sufficient. $\ \mathcal{L}_{\mathrm{basic}}$ thus represents a
generalized Wess-Zumino-Witten term \cite%
{Wess:1971yu,Witten:1983tw,D'Hoker:1994ti,D'Hoker:1995it}. The corresponding
action can be written as an integral of closed $GAL\left( D,1\right) $%
-invariant $D+1$ form $\omega _{D+1}$ (which is however not an exterior
derivative of any $GAL\left( D,1\right) $-invariant $D$-form). The
integration is taken over $D+1$ dimensional ball $B_{D+1}$ whose boundary is
the compactified space-time $S_{D}=\partial B_{D+1}$

\begin{equation}
S^{WZW}=\int_{B_{D+1}}\omega _{D+1}.  \label{S_WZW}
\end{equation}%
The basis\footnote{%
I.e. the basis of the cohomology $H^{D+1}\left( GAL(D,1)/SO(1,D-1),\mathbb{R}%
\right) $; see \cite{Goon:2012dy} for more detail.} of such forms was found
in \cite{Goon:2012dy} and has $D+1$ elements $\omega _{D+1}^{\left( n\right)
}$, explicitly%
\begin{equation}
\omega _{D+1}^{(n)}=\mathrm{d}\beta _{d}^{(n)}=\varepsilon _{\mu _{1}\ldots
\mu _{D}}\omega _{A}\wedge \omega _{B}^{\mu _{1}}\wedge \ldots \wedge \omega
_{B}^{\mu _{n-1}}\wedge \omega _{P}^{\mu _{n}}\wedge \ldots \wedge \omega
_{P}^{\mu _{D}},  \label{omega_n_D+1}
\end{equation}%
where the $GAL\left( D,1\right) $-non-invariant $D$-forms $\beta _{D}^{(n)}$
are%
\begin{eqnarray}
\beta _{D}^{(n)} &=&\varepsilon _{\mu _{1}\ldots \mu _{D}}\phi ~\mathrm{d}%
L^{\mu _{1}}\wedge \ldots \wedge \mathrm{d}L^{\mu _{n-1}}\wedge \mathrm{d}%
x^{\mu _{n}}\wedge \ldots \wedge \mathrm{d}x^{\mu _{D}}  \notag \\
&&+\frac{n-1}{2(D-n+2)}\varepsilon _{\mu _{1}\ldots \mu _{D}}L^{2}\mathrm{d}%
L^{\mu _{1}}\wedge \ldots \wedge \mathrm{d}L^{\mu _{n-2}}\wedge \mathrm{d}%
x^{\mu _{n-1}}\wedge \ldots \wedge \mathrm{d}x^{\mu _{D}}.
\end{eqnarray}%
We can thus write according to the Stokes theorem%
\begin{equation}
\int_{B_{D+1}}\omega _{D+1}^{(n)}=\int_{\partial B_{D+1}}\beta
_{D}^{(n)}=\int_{S_{D}}\beta _{D}^{(n)}
\end{equation}%
As the last step we impose the IHC constraint (\ref{IHC}) in the last
formula. As a result we get%
\begin{equation}
\int_{S_{D}}\beta _{D}^{(n)}|_{IHC}=\frac{1}{n}\int_{S_{D}}\mathrm{d}^{D}x%
\mathcal{L}_{n} ,  \label{identifikaceLn}
\end{equation}%
where%
\begin{eqnarray}
\mathcal{L}_{n} &=&\phi \varepsilon ^{\mu _{1}\ldots \mu _{D}}\varepsilon
^{\nu _{1}\ldots \nu _{D}}\prod\limits_{i=1}^{n-1}\partial _{\mu
_{i}}\partial _{\nu _{i}}\phi \prod\limits_{j=n}^{D}\eta _{\mu _{j}\nu _{j}}
\notag \\
&=&\left( -1\right) ^{D-1}(D-n+1)!\phi \det \left\{ \partial ^{\nu
_{i}}\partial _{\nu _{j}}\phi \right\} _{i,j=1}^{n-1}.  \label{L_n}
\end{eqnarray}%
$\mathcal{L}_{n}$ is up to a constant one of the traditional forms of the Galileon Lagrangian.
Here we use the convention $\eta =\mathrm{diag}\left( 1,-1,\ldots ,-1\right)
$ and $\varepsilon ^{01\ldots D-1}=1$. The most general basic Galileon
action can be therefore written in the form%
\begin{equation}
S_{\mathrm{basic}}=\sum_{n=1}^{D+1}nd_{n}\int_{B_{D+1}}\omega
_{D+1}^{(n)}|_{IHC}=\int_{S_{D}}\mathrm{d}^{D}x\sum_{n=1}^{D+1}d_{n}\mathcal{%
L}_{n}.  \label{S_basic}
\end{equation}%
where $d_{n}$ are real constants.

\section{Coset transformations and Galileon dualities\label{coset and
dualities}}

In the previous section the coset space $GAL\left( d,1\right) /SO(1,d-1)$
has been parametrized by means of the coordinates $x^{\mu }$, $L^{\mu }$ and
$\phi $ according to the choice $U$ of the representative of each coset,
where
\begin{equation}
U=\mathrm{e}^{\mathrm{i}P\cdot x}\mathrm{e}^{\mathrm{i}A\phi +\mathrm{i}%
B\cdot L},
\end{equation}%
and then the redundant field $L^{\mu }$ has been fixed by means of imposing
the IHC constraint%
\begin{equation}
\omega _{A}=\mathrm{d}\phi -L\cdot \mathrm{d}x=0.
\end{equation}%
Therefore any transformation of the coset space which preserves the IHC
constraint defines a consistent transformation of the Galileon field. More
formally, suppose that we have a general transformation on the coset space
which is expressed in terms of the coordinates $x^{\mu }$, $L^{\mu }$ and $%
\phi $ as%
\begin{eqnarray}
x^{\prime \mu } &=&\xi ^{\mu }\left( x,L,\phi \right)  \notag \\
L^{\prime \mu } &=&\Lambda ^{\mu }\left( x,L,\phi \right)  \notag \\
\phi ^{\prime } &=&f\left( x,L,\phi \right) .  \label{coset_transformations}
\end{eqnarray}%
Provided the IHC  is preserved by this transformation, i.e. when
the following implication holds%
\begin{equation}
\omega _{A}=0\Rightarrow \omega _{A}^{^{\prime }}=0,  \label{consistency}
\end{equation}%
where%
\begin{equation}
\omega _{A}^{^{\prime }}\equiv \mathrm{d}\phi ^{^{\prime }}-L^{^{\prime
}}\cdot \mathrm{d}x^{^{\prime }},  \label{IHC_transformed}
\end{equation}%
then a well defined transformation of the Galileon field can be obtained as
\begin{eqnarray}
x^{^{\prime }\mu } &=&\xi ^{\mu }\left( x,\partial \phi \left( x\right)
,\phi \left( x\right) \right)  \notag \\
\phi ^{^{\prime }}(x^{\prime }) &=&f\left( x,\partial \phi \left( x\right)
,\phi \left( x\right) \right) .
\end{eqnarray}%
As a consequence of $\omega _{A}^{^{\prime }}=0$ we get a consistent
relation for the transformation of the derivatives%
\begin{equation}
\partial ^{^{\prime }\mu }\phi ^{\prime }\left( x^{\prime }\right) =\Lambda
^{\mu }\left( x,\partial \phi \left( x\right) ,\phi \left( x\right) \right) .
\end{equation}%
An important class of such transformations are dualities, i.e. those
transformations for which the general basic Galileon action is
form-invariant. This means that, provided%
\begin{equation}
S_{\mathrm{basic}}\left[ \phi \right] =\int \mathrm{d}^{D}x%
\sum_{n=1}^{D+1}d_{n}\mathcal{L}_{n}\left[ \phi \right]
\end{equation}%
with some set $\left\{ d_{n}\right\} _{n=1}^{D+1}$ of the couplings, then
\begin{equation}
S_{\mathrm{basic}}\left[ \phi ^{\prime }\right] =\int \mathrm{d}%
^{D}x\sum_{n=1}^{D+1}d_{n}^{\prime }\mathcal{L}_{n}\left[ \phi \right]
\end{equation}%
with new set $\left\{ d_{n}^{\prime }\right\} _{n=1}^{D+1}$. In the case
when $\left\{ d_{n}\right\} _{n=1}^{D+1}=\left\{ d_{n}^{\prime }\right\}
_{n=1}^{D+1}$ the duality becomes a symmetry of the basic Galileon action.

A large class of such dualities has been classified in \cite{Kampf:2014rka}.
These dualites can be identified with group of matrices $\boldsymbol{M}\in
GL\left( 2, \mathbb{R} \right) $ which act on the coset space as follows.
The transformation of the coordinated $x^{\mu }$ and $L^{\mu }$ are given by
the matrix multiplication
\begin{equation}
\left(
\begin{array}{c}
x^{\prime \mu } \\
L^{\prime \mu }%
\end{array}%
\right) =\boldsymbol{M}\left(
\begin{array}{c}
x^{\mu } \\
L^{\mu }%
\end{array}%
\right)  \label{x_L_transformation}
\end{equation}%
while the transformation of $\phi $ can be written in a compact form as%
\begin{equation}
\phi ^{\prime }-\frac{1}{2}L^{\prime }\cdot x^{\prime }=\left( \phi -\frac{1%
}{2}L\cdot x\right) \det \boldsymbol{M}.  \label{phi_transformation}
\end{equation}%
The new set of couplings is then a linear combination of the original ones%
\begin{equation}
d_{n}^{\prime }=\sum_{m=1}^{D+1}A_{nm}\left( \boldsymbol{M}\right) d_{m}
\end{equation}%
for appropriate matrix $A_{nm}\left( \boldsymbol{M}\right) $. Composition of
the transformations (\ref{x_L_transformation}), (\ref{phi_transformation})
is in one-to-one corresponcence with the matrix multiplication in $GL\left(
2, \mathbb{R} \right) $ and the theory space (i.e. the $(D+1)$ dimensional
vector space with coordinates $d_{m}$) caries its linear representation (see
\cite{Kampf:2014rka} for more details). Let us note that for special choice
of the matrix $\boldsymbol{M}=\alpha _{D}\left( \theta \right) $ where%
\begin{equation}
\alpha _{D}\left( \theta \right) =\left(
\begin{array}{cc}
1 & -2\theta \\
0 & 1%
\end{array}%
\right)  \label{alpha_D_matrix}
\end{equation}%
we recover a one parametric subgroup of dualities%
\begin{equation}
x^{\prime }=x-2\theta \partial \phi \left( x\right) ,~~~\phi ^{\prime
}\left( x^{\prime }\right) =\phi \left( x\right) -\theta \partial \phi
\left( x\right) \cdot \partial \phi \left( x\right)  \label{old_duality}
\end{equation}%
which has been discussed in \cite{deRham:2013hsa}, \cite{deRham:2014lqa} and
\cite{Creminelli:2014zxa}. Let us remind, that the duality (\ref{old_duality}%
) preserves the on-shell $S-$matrix, i.e. the theories related with this
transformation describe the same on-shell physics (see \cite{Kampf:2014rka}
for more detail).

\section{Special Galileon duality}

In this section we introduce another very special type of the coset space
transformations (\ref{coset_transformations}) satisfying (\ref{consistency}%
). We will show that it is possible to find a new set of duality (and even
symmetry) transformations outside the class (\ref{x_L_transformation}), (\ref%
{phi_transformation}) provided we restrict ourselves to a special two (or
three)-parametric families of the basic Galileon actions. Within the new set
of dualities we will find a subset corresponding to the symmetry discussed
by Hinterbichler and Joyce in \cite{Hinterbichler:2015pqa}. As a particular
member of the family of actions mentioned above we will identify a theory
known as Special Galileon.

In what follows we will distinguish between two branches of the
abovementioned duality transformation. The most natural identification of
these two branches is in terms of the appropriately choosen set of complex
and real coordinates on the coset space.

\subsection{Complex coordinates}

For further convenience we introduce first the following complex
combinations of the coset coordinates%
\begin{equation}
Z=x+\frac{\mathrm{i}}{\alpha }L,~~~~~\overline{Z}=x-\frac{\mathrm{i}}{\alpha
}L ,  \label{ZZbar}
\end{equation}%
where $\alpha $ is a real parameter with canonical dimension $\dim \alpha
=\left( D+2\right) /2$. Note that the building blocks $\omega _{A}$, $\omega
_{B}$ and $\omega _{P}$ (see (\ref{forms})) read in these coordinates%
\begin{eqnarray}
\omega _{A} &=&\mathrm{d}\phi +\mathrm{i}\frac{\alpha }{4}\left( Z-\overline{%
Z}\right) \cdot \left( \mathrm{d}Z+\mathrm{d}\overline{Z}\right) \\
\omega _{P} &=&\frac{1}{2}\left( \mathrm{d}Z+\mathrm{d}\overline{Z}\right) \\
\omega _{B} &=&-\frac{\mathrm{i}\alpha }{2}\left( \mathrm{d}Z-\mathrm{d}%
\overline{Z}\right) .
\end{eqnarray}%
Let now $G^{\mu \nu }=G^{\nu \mu \text{ }}$be constant symmetric real tensor
and $\theta $ be a real number which will play a role of the parameter of
the transformation (here we assume $\dim G^{\mu \nu }=0$ and $\dim \theta
=-\dim \alpha $). Let us denote $U\left( \theta \right) $ the following
matrix
\begin{equation}
U\left( \theta \right) =\exp \left( -\mathrm{i}\alpha \theta \mathcal{G}%
\right),  \label{U_matrix}
\end{equation}%
where in the matrix notation
\begin{equation}
\left( \mathcal{G}\right) _{~\nu }^{\mu }=G^{\mu \alpha }\eta _{\alpha \nu
}=G_{\nu }^{\mu } ,  \label{G_matrix}
\end{equation}%
and thus%
\begin{equation}
U\left( \theta \right) _{~\nu }^{\mu }=\delta _{\nu }^{\mu }-\mathrm{i}%
\alpha \theta G_{\nu }^{\mu }-\frac{1}{2}\alpha ^{2}\theta ^{2}G_{\alpha
}^{\mu }G_{\nu }^{\alpha }+\ldots
\end{equation}%
Let us introduce the following transformation of the coset coordinates%
\footnote{%
Note that the transformation of $\overline{Z}$ is consistent with the
relation $Z^{\ast }=\overline{Z}$.}%
\begin{eqnarray}
Z_{\theta } &=&U\left( \theta \right) \cdot Z  \notag \\
\overline{Z}_{\theta } &=&U\left( -\theta \right) \cdot \overline{Z}  \notag
\\
\phi _{\theta } &=&\phi +\mathrm{i}\frac{\alpha }{8}\left( Z^{2}-\overline{Z}%
^{2}\right) -\mathrm{i}\frac{\alpha }{8}\left( Z_{\theta }^{2}-\overline{Z}%
_{\theta }^{2}\right)  \label{hidden symmetry}
\end{eqnarray}%
As a result of the transformation we get%
\begin{eqnarray}
\mathrm{d}Z_{\theta } &=&U\left( \theta \right) \cdot \mathrm{d}Z  \notag \\
\mathrm{d}\overline{Z}_{\theta } &=&U\left( -\theta \right) \cdot \mathrm{d}%
\overline{Z}.  \label{differentials}
\end{eqnarray}%
Using the properties of the matrix $U\left( \theta \right) $ it is then an
easy exercise (see Appendix \ref{IHC1}) to show that the form $\omega _{A}$
is invariant, i.e.
\begin{eqnarray*}
\left[ \omega _{A}\right] _{\theta } &\equiv &\mathrm{d}\phi _{\theta }+%
\mathrm{i}\frac{\alpha }{4}\left( Z_{\theta }-\overline{Z}_{\theta }\right)
\cdot \left( \mathrm{d}Z_{\theta }+\mathrm{d}\overline{Z}_{\theta }\right) \\
&=&\mathrm{d}\phi +\mathrm{i}\frac{\alpha }{4}\left( Z-\overline{Z}\right)
\cdot \left( \mathrm{d}Z+\mathrm{d}\overline{Z}\right) =\omega _{A}.
\end{eqnarray*}%
Thus the transformation (\ref{hidden symmetry}) respects the IHC constraint
in the sense discussed above and therefore induces a well defined
transformation of the Galileon field. Let us note that this definition
guaranties that the combination%
\begin{equation}
\phi +\mathrm{i}\frac{\alpha }{8}\left( Z^{2}-\overline{Z}^{2}\right) =\phi -%
\frac{1}{2}L\cdot x  \label{invariant}
\end{equation}%
is invariant with respect to the transformation (\ref{hidden symmetry}) (cf.
also (\ref{phi_transformation})).

Let us now construct a basic Galileon action with nice transformation
properties with respect to the transformation (\ref{hidden symmetry}). As we
have discussed above (see (\ref{S_basic})), any such action is a linear
combination of the integrals of the basic forms $\omega _{D+1}^{\left(
n+1\right) }$%
\begin{equation}
\omega _{D+1}^{\left( n+1\right) }=\varepsilon _{\mu _{1}\mu _{2}\ldots \mu
_{D}}\omega _{A}\wedge \omega _{B}^{\mu _{1}}\wedge \ldots \wedge \omega
_{B}^{\mu _{n}}\wedge \omega _{P}^{\mu _{n+1}}\wedge \ldots \wedge \omega
_{P}^{\mu _{D}}
\end{equation}%
over $D+1$ dimensional ball $B_{D+1}$. Because the basic building blocks $%
\omega _{P}$ and $\omega _{B}$ are linear combinations of $\mathrm{d}Z$ and $%
\mathrm{d}\overline{Z}$ (see (\ref{forms})), it is natural to consider the
following form%
\begin{eqnarray}
\Omega &=&\varepsilon _{\mu _{1}\mu _{2}\ldots \mu _{D}}\omega _{A}\wedge
\mathrm{d}Z^{\mu _{1}}\wedge \mathrm{d}Z^{\mu _{2}}\ldots \wedge \mathrm{d}%
Z^{\mu _{D}}  \notag \\
&=&\sum_{n=0}^{D}\left(
\begin{array}{c}
D \\
n%
\end{array}%
\right) \left( \frac{\mathrm{i}}{\alpha }\right) ^{n}\omega _{D+1}^{\left(
n+1\right) }.  \label{Omega}
\end{eqnarray}%
The latter transforms under (\ref{hidden symmetry}) as follows%
\begin{eqnarray}
\Omega _{\theta } &=&\varepsilon _{\mu _{1}\mu _{2}\ldots \mu _{D}}\left[
\omega _{A}\right] _{\theta }\wedge \mathrm{d}Z_{\theta }^{\mu _{1}}\wedge
\mathrm{d}Z_{\theta }^{\mu _{2}}\ldots \wedge \mathrm{d}Z_{\theta }^{\mu
_{D}}  \notag \\
&=&\varepsilon _{\mu _{1}\mu _{2}\ldots \mu _{D}}U\left( \theta \right)
_{\nu _{1}}^{\mu _{1}}\ldots U\left( \theta \right) _{\nu _{D}}^{\mu
_{D}}\omega _{A}\wedge \mathrm{d}Z^{\nu _{1}}\wedge \ldots \wedge \mathrm{d}%
Z^{\nu _{D}}  \notag \\
&=&\det U\left( \theta \right) \varepsilon _{\nu _{1}\nu _{2}\ldots \nu
_{D}}\omega _{A}\wedge \mathrm{d}Z^{\nu _{1}}\wedge \ldots \wedge \mathrm{d}%
Z^{\nu _{D}},
\end{eqnarray}%
and thus%
\begin{equation}
\Omega _{\theta }=\det U\left( \theta \right) \Omega =\mathrm{e}^{-\mathrm{i}%
\alpha \theta \mathrm{tr}\mathcal{G}}\Omega .  \label{Omega_theta}
\end{equation}%
In the same way, the form%
\begin{equation}
\overline{\Omega }=\varepsilon _{\mu _{1}\mu _{2}\ldots \mu _{D}}\omega
_{A}\wedge \mathrm{d}\overline{Z}^{\mu _{1}}\wedge \mathrm{d}\overline{Z}%
^{\mu _{2}}\ldots \wedge \mathrm{d}\overline{Z}^{\mu _{D}}  \label{Omega_bar}
\end{equation}%
transforms as%
\begin{equation}
\overline{\Omega }_{\theta }=\det U\left( -\theta \right) \overline{\Omega }=%
\mathrm{e}^{\mathrm{i}\alpha \theta \mathrm{tr}\mathcal{G}}\Omega.
\label{Omega_bar_theta}
\end{equation}%
This results suggest to construct a two-parametric family of basic Galileon
actions\footnote{%
The prefactor $\alpha /2$ avoids the $\alpha $-dependence of the kinetic
term.}
\begin{equation}
S\left( \alpha ,\beta \right) =\frac{1}{2\mathrm{i}}\alpha
\int_{B_{D+1}}\left( \mathrm{e}^{\mathrm{i}\beta }\Omega -\mathrm{e}^{-%
\mathrm{i}\beta }\overline{\Omega }\right) |_{IHC} ,  \label{action_alpha_xi}
\end{equation}%
where $\beta $ is a real parameter. Such actions have (as a consequence of (%
\ref{Omega_theta}) and (\ref{Omega_bar_theta})) very simple transformation
property with respect to (\ref{hidden symmetry}),
\begin{eqnarray}
S_{\theta }\left( \alpha ,\beta \right) &=&\frac{1}{2\mathrm{i}}\alpha
\int_{B_{D+1}}\left( \mathrm{e}^{\mathrm{i}\beta }\Omega _{\theta }-\mathrm{e%
}^{-\mathrm{i}\beta }\overline{\Omega }_{\theta }\right) |_{IHC} \\
&=&\frac{1}{2\mathrm{i}}\alpha \int_{B_{D+1}}\left( \mathrm{e}^{\mathrm{i}%
\beta }\mathrm{e}^{-\mathrm{i}\alpha \theta \mathrm{tr}\mathcal{G}}\Omega -%
\mathrm{e}^{-\mathrm{i}\beta }\mathrm{e}^{\mathrm{i}\alpha \theta \mathrm{tr}%
\mathcal{G}}\overline{\Omega }\right) |_{IHC},
\end{eqnarray}%
i.e. the parameter $\beta $ is shifted according to%
\begin{equation}
S_{\theta }\left( \alpha ,\beta \right) =S\left( \alpha ,\beta -\alpha
\theta G_{\mu }^{\mu }\right) .
\end{equation}%
The transformation (\ref{hidden symmetry}) is therefore a duality
transformation and for traceless tensor $G^{\mu \nu }$ it is a symmetry of
the two parameter family of actions $S\left( \alpha ,\beta \right) $. In the
traditional notation we get using (\ref{S_basic}) and (\ref{Omega})%
\begin{equation}
S\left( \alpha ,\beta \right) =\int \mathrm{d}^{D}x\mathcal{L}\left( \alpha
,\beta \right),
\end{equation}%
where%
\begin{equation}
\mathcal{L}\left( \alpha ,\beta \right) =\sum_{n=1}^{D+1}d_{n}\left( \alpha
,\beta \right) \mathcal{L}_{n}  \label{S_real}
\end{equation}%
with $\mathcal{L}_{n}$ given by (\ref{L_n}) and where the couplings $%
d_{n}\left( \alpha ,\beta \right) $ read explicitly
\begin{eqnarray}
d_{2n}\left( \alpha ,\beta \right) &=&\frac{\left( -1\right) ^{n}}{2n}\left(
\begin{array}{c}
D \\
2n-1%
\end{array}%
\right) \frac{\cos \beta }{\alpha ^{2\left( n-1\right) }}, \\
d_{2n+1}\left( \alpha ,\beta \right) &=&\frac{\left( -1\right) ^{n}}{2n+1}%
\left(
\begin{array}{c}
D \\
2n%
\end{array}%
\right) \frac{\sin \beta }{\alpha ^{2n-1}}.
\end{eqnarray}%
Especially for traceless tensor $G^{\mu \nu }$ we get invariance of the
Lagrangian $\mathcal{L}\left( \alpha ,\beta \right) $ under (\ref{hidden
symmetry}) (with IHC  imposed) up to a total derivative.

The infinitesimal form of the duality transformation reads%
\begin{eqnarray}
Z_{\theta }^{\mu } &=&Z^{\mu }-\mathrm{i}\alpha \theta G_{\nu }^{\mu }Z^{\nu
}  \notag \\
\overline{Z}_{\theta }^{\mu } &=&\overline{Z}^{\mu }+\mathrm{i}\alpha \theta
G_{\nu }^{\mu }\overline{Z}^{\nu }  \notag \\
\phi _{\theta } &=&\phi -\frac{\alpha ^{2}}{4}\theta \left( Z^{\mu }G_{\mu
\nu }Z^{\nu }+\overline{Z}^{\mu }G_{\mu \nu }\overline{Z}^{\nu }\right),
\end{eqnarray}%
or using the IHC %
\begin{align}
x_{\theta }^{\mu }& =x^{\mu }+\theta G^{\mu \nu }\partial _{\nu }\phi  \notag
\\
\phi _{\theta }\left( x_{\theta }\right) & =\phi \left( x\right) -\frac{%
\alpha ^{2}}{2}\theta G^{\mu \nu }\left( x_{\mu }x_{\nu }-\frac{1}{\alpha
^{2}}\partial _{\mu }\phi \left( x\right) \partial _{\nu }\phi \left(
x\right) \right).  \label{infinitesimal}
\end{align}%
Equivalently we can write%
\begin{equation}
\phi _{\theta }\left( x\right) =\phi \left( x\right) -\frac{\theta }{2}%
G^{\mu \nu }\left( \alpha ^{2}x_{\mu }x_{\nu }+\partial _{\mu }\phi \left(
x\right) \partial _{\nu }\phi \left( x\right) \right) .
\label{infinitesimal1}
\end{equation}%
In the latter formula we can recognize a generalization of the hidden
Galileon symmetry of Hinterbichler and Joyce \cite{Hinterbichler:2015pqa}.
The theory known as a Special Galileon corresponds then (up to an overall
normalization) to particular value of parameter $\beta $, namely for $\beta
=k\pi ,~k\in \mathbb{Z} $. Therefore the hidden Galileon symmetry is a
special case of more general \textquotedblleft hidden Galileon
duality\textquotedblleft\ (\ref{hidden symmetry}) for traceless $G^{\mu \nu
} $ and there is in fact a two parameter family of Lagrangians invariant
with respect to this symmetry.

\subsection{Real coordinates}

We can repeat the above consideration also for imaginary parameter $\alpha =%
\mathrm{i}a$. Let us introduce real coordinates on the coset space according
to%
\begin{equation}
Z^{\pm }=x\pm \frac{1}{a}L,  \label{Zpm}
\end{equation}%
and define the transformation matrix $U\left( \pm \theta \right) $ as
\begin{equation*}
U\left( \pm \theta \right) =\exp \left( \pm a\theta \mathcal{G}\right) .
\end{equation*}%
Here $\mathcal{G}$ is defined as above (\ref{G_matrix}). The transformation
of the coset coordinates is then%
\begin{eqnarray}
Z_{\theta }^{\pm } &=&U\left( \pm \theta \right) \cdot Z^{\pm }  \notag \\
\phi _{\theta } &=&\phi -\frac{a}{8}\left( Z^{+2}-Z^{-2}\right) +\frac{a}{8}%
\left( Z_{\theta }^{+2}-Z_{\theta }^{-2}\right) ,  \label{duality_imaginary}
\end{eqnarray}%
and for the forms $\omega _{A}$ and $\mathrm{d}Z^{\pm }$ we get following
transformation rules%
\begin{equation}
\left[ \omega _{A}\right] _{\theta }=\omega _{A},~~~~\mathrm{d}Z_{\theta
}^{\pm }=U\left( \pm \theta \right) \cdot \mathrm{d}Z^{\pm }.
\end{equation}%
In analogy with (\ref{Omega}) and (\ref{Omega_bar}) we can now construct two
real forms%
\begin{eqnarray}
\Omega ^{\pm } &=&\varepsilon _{\mu _{1}\mu _{2}\ldots \mu _{D}}\omega
_{A}\wedge \mathrm{d}Z^{\pm \mu _{1}}\wedge \mathrm{d}Z^{\pm \mu _{2}}\ldots
\wedge \mathrm{d}Z^{\pm \mu _{D}}  \notag \\
&=&\sum_{n=0}^{D}\left(
\begin{array}{c}
D \\
n%
\end{array}%
\right) \left( \pm \frac{1}{a}\right) ^{n}\omega _{D+1}^{\left( n+1\right) },
\label{omega_plus_minus}
\end{eqnarray}%
transforming as%
\begin{equation}
\Omega _{\theta }^{\pm }=\Omega ^{\pm }\det \left[ \exp \left( \pm a\theta
\mathrm{tr}\mathcal{G}\right) \right] =\mathrm{e}^{\pm a\theta G_{\mu }^{\mu
}}\Omega ^{\pm }.  \label{omega+-transformation}
\end{equation}%
Let us therefore assume the following actions
\begin{equation}
S_{\pm }\left( a,c_{\pm }\right) =\pm \frac{1}{2}a\mathrm{e}^{c_{\pm
}}\int_{B_{D+1}}\Omega ^{\pm }|_{IHC}=\int \mathrm{d}^{D}x%
\sum_{n=1}^{D+1}d_{n}^{\pm }\left( a,c_{\pm }\right) \mathcal{L}_{n},
\label{s_plus_minus}
\end{equation}%
where%
\begin{equation}
d_{n}^{\pm }\left( a,c_{\pm }\right) =\frac{1}{n}\mathrm{e}^{c_{\pm }}\left(
\begin{array}{c}
D \\
n-1%
\end{array}%
\right) \frac{\left( -1\right) ^{n}}{a^{n-2}}.
\end{equation}%
According to (\ref{omega+-transformation}), these actions are related by
duality (\ref{duality_imaginary})%
\begin{equation}
S_{\pm }\left( a,c_{\pm }\right) _{\theta }=S_{\pm }\left( a,c_{\pm }\pm
a\theta G_{\mu }^{\mu }\right) ,
\end{equation}%
which becomes a symmetry of $S_{\pm }\left( a,c_{\pm }\right) $ for
traceless $G^{\mu \nu }$.

However, the theories corresponding either to only $S_{+}\left(
a,c_{+}\right) $ or to only $S_{-}\left( a,c_{-}\right) $ are in some sense
trivial. As it is shown in Appendix \ref{duality_on_special_galileon}, under
the $GL\left( 2, \mathbb{R} \right) $ duality transformation (\ref%
{x_L_transformation}), (\ref{phi_transformation}) with matrix $\boldsymbol{M}%
=\alpha _{D}\left( \sigma \right) $ (see (\ref{alpha_D_matrix})) we
transform $d_{n}^{\pm }\left( a,c_{\pm }\right) $ into%
\begin{equation}
d_{n}^{\pm }\left( a,c_{\pm }\right) _{\sigma }=\frac{1}{2n}a\mathrm{e}%
^{c_{\pm }}\left(
\begin{array}{c}
D \\
n-1%
\end{array}%
\right) \left( \pm \frac{1}{a}-2\sigma \right) ^{n-1}.  \label{d_pm_sigma}
\end{equation}%
Thus for $\boldsymbol{M}_{\pm }=\alpha _{D}\left( \pm 1/2\alpha \right) $,
the actions $S_{\pm }\left( a,c_{\pm }\right) $ are dual to the theory with
action%
\begin{equation}
S_{1}\left( a,c_{\pm }\right) =\frac{\left( -1\right) ^{D-1}D!}{2}a\mathrm{e}%
^{c_{\pm }}\int \mathrm{d}^{D}x\phi .
\end{equation}%
To get a nontrivial theory we have therefore to construct a linear
combination of both actions $S_{\pm }\left( a,c_{\pm }\right) $ with nonzero
coefficients,
\begin{equation}
S\left( a,c_{+},c_{-}\right) =\frac{1}{2}a\int_{B_{D+1}}\left( \mathrm{e}%
^{c_{+}}\Omega ^{+}-\mathrm{e}^{c_{-}}\Omega ^{-}\right) |_{IHC}=\int
\mathrm{d}^{D}x\sum_{n=1}^{D+1}d_{n}^{\pm }\left( a,c_{+},c_{-}\right)
\mathcal{L}_{n},
\end{equation}%
where%
\begin{equation}
d_{n}^{\pm }\left( a,c_{+},c_{-}\right) =\frac{1}{2n}\left(
\begin{array}{c}
D \\
n-1%
\end{array}%
\right) \frac{1}{a^{n-2}}\left( \mathrm{e}^{c_{+}}+\left( -1\right) ^{n}%
\mathrm{e}^{c_{-}}\right) .
\end{equation}%
Under (\ref{duality_imaginary}) $S_{\pm }\left( a,c_{+},c_{-}\right) $
transforms according to%
\begin{equation}
S\left( a,c_{+},c_{-}\right) _{\theta }=S_{\pm }\left( a,c_{+}+a\theta
G_{\mu }^{\mu },c_{-}-a\theta G_{\mu }^{\mu }\right) ,  \label{Scpcm}
\end{equation}%
and again, the case of traceless $G_{\mu }^{\mu }=0$ yields a symmetry of
the whole family $S\left( a,c_{+},c_{-}\right) $. Note that the special case
$S_{\pm }\left( a,b,-b\right) $ is an analytic continuation of the action $%
S\left( \alpha ,\beta \right) $ (cf. (\ref{action_alpha_xi})) to imaginary
values of the parameters, namely $\alpha =\mathrm{i}a$ and $\beta =-\mathrm{i%
}b$
\begin{equation}
S\left( \mathrm{i}a,-\mathrm{i}b\right) =\frac{1}{2}a\int_{B_{D+1}}\left(
\mathrm{e}^{b}\Omega ^{+}-\mathrm{e}^{-b}\Omega ^{-}\right) |_{IHC}=\int
\mathrm{d}^{D}x\sum_{n=1}^{D+1}d_{n}\left( \mathrm{i}a,-\mathrm{i}b\right)
\mathcal{L}_{n},
\end{equation}%
where
\begin{eqnarray}
d_{2n}\left( \mathrm{i}a,-\mathrm{i}b\right) &=&\frac{1}{2n}\left(
\begin{array}{c}
D \\
2n-1%
\end{array}%
\right) \frac{\cosh b}{a^{2\left( n-1\right) }}  \notag \\
d_{2n+1}\left( \mathrm{i}a,-\mathrm{i}b\right) &=&\frac{1}{2n+1}\left(
\begin{array}{c}
D \\
2n%
\end{array}%
\right) \frac{\sinh b}{a^{2n-1}}.
\end{eqnarray}%
Under (\ref{duality_imaginary}) this actions transform as%
\begin{equation}
S_{\theta }\left( \mathrm{i}a,-\mathrm{i}b\right) =S\left( \mathrm{i}a,-%
\mathrm{i}b-\mathrm{i}a\theta G_{\mu }^{\mu }\right) .
\end{equation}

The infinitesimal transformation (\ref{duality_imaginary}) reads in terms of
$x$ and $\phi $ (after using the IHC)%
\begin{eqnarray}
x_{\theta }^{\mu } &=&x^{\mu }+\theta G^{\mu \nu }\partial _{\nu }\phi
\notag \\
\phi _{\theta }\left( x_{\theta }\right) &=&\phi \left( x\right) +\frac{a^{2}%
}{2}\theta G^{\mu \nu }\left( x_{\mu }x_{\nu }+\frac{1}{a^{2}}\partial _{\mu
}\phi \left( x\right) \partial _{\nu }\phi \left( x\right) \right)
\end{eqnarray}%
or finally%
\begin{equation}
\phi _{\theta }\left( x\right) =\phi \left( x\right) +\frac{\theta }{2}%
G^{\mu \nu }\left( a^{2}x_{\mu }x_{\nu }-\partial _{\mu }\phi \left(
x\right) \partial _{\nu }\phi \left( x\right) \right)
\end{equation}%
This corresponds to the second branch of the hidden Galileon symmetry of
Hinterbichler and Joyce \cite{Hinterbichler:2015pqa}. The corresponding
Special Galileon is a particular case of $S\left( a,c_{+},c_{-}\right) $
with $c_{+}=c_{-}$.

\subsection{Special Galileon duality and $GL\left( 2, \mathbb{R} \right) $
duality}

Let us add a note on the interrelation of the above Special Galileon duality
and the $GL\left( 2, \mathbb{R} \right) $ duality (\ref{x_L_transformation}%
), (\ref{phi_transformation}) discussed in Section \ref{coset and dualities}%
. For $G^{\mu \nu }=\eta ^{\mu \nu }$ the transformations (\ref{hidden
symmetry}) and (\ref{duality_imaginary}) correspond to the dualities (\ref%
{x_L_transformation}), (\ref{phi_transformation}) with $SL\left( 2, \mathbb{R%
} \right) $ matrices $\boldsymbol{M}_{\alpha }\left( \theta \right) $ and $%
\boldsymbol{M}_{a}\left( \theta \right) $ respectively, namely%
\begin{eqnarray}
\boldsymbol{M}_{\alpha }\left( \theta \right) &=&\left(
\begin{array}{cc}
\cos \alpha \theta & \alpha ^{-1}\sin \alpha \theta \\
-\alpha \sin \alpha \theta & \cos \alpha \theta%
\end{array}%
\right) \\
\boldsymbol{M}_{a}\left( \theta \right) &=&\left(
\begin{array}{cc}
\cosh a\theta & a^{-1}\sinh a\theta \\
a\sinh a\theta & \cosh a\theta%
\end{array}%
\right) .
\end{eqnarray}%
The $G^{\mu \nu }=\eta ^{\mu \nu }$ case is therefore not only duality of
the Special Galileon but also a duality of the most general Galileon action.
Note that in the limit $\alpha ,a\rightarrow 0$ we get%
\begin{equation}
\boldsymbol{M}_{\alpha ,a}\left( \theta \right) \overset{\alpha
,a\rightarrow 0}{\rightarrow }\left(
\begin{array}{cc}
1 & \theta \\
0 & 1%
\end{array}%
\right) =\alpha _{D}\left( -\theta /2\right)
\end{equation}%
and the resulting matrix corresponds to the one parametric subgroup of
dualities (\ref{alpha_D_matrix}), (\ref{old_duality}).

\subsection{Possible generalization of Special Galileon duality}

In this subsection we will discuss one possible generalization of the
Special Galileon duality. Originally this kind of transformation has been
discussed by Noller, Sivanesan and von Strauss in \cite{Noller:2015rea}.
Though this transformation is apparently a nontrivial symmetry of a two
parameter family of Galileon actions, it can be shown, that such a family is
in some sense trivial being dual (with respect to the duality (\ref%
{x_L_transformation}), (\ref{phi_transformation})) to the free theory with
tadpole term. Here we give a coset space formulation of this generalized
symmetry and construct the invariant actions as the appropriate manifestly
invariant Wess-Zumino terms (\ref{S_WZW}).

Let us try to generalize the duality (\ref{duality_imaginary}) and assume
the following coordinate transformation on the coset space%
\begin{eqnarray}
Z_{\theta }^{+\mu } &=&Z^{+\mu }  \notag \\
Z_{\theta }^{-\mu } &=&Z^{-\mu }+\frac{2\theta }{a}G^{\mu \mu _{1}\mu
_{2}\ldots \mu _{N-1}}Z_{\mu _{1}}^{+}Z_{\mu _{2}}^{+}\ldots Z_{\mu
_{N-1}}^{+}  \notag \\
\phi _{\theta } &=&\phi +\frac{a}{8}\left( Z^{+}-Z^{-}\right) ^{2}-\frac{a}{8%
}\left( Z_{\theta }^{+}-Z_{\theta }^{-}\right) ^{2}-\frac{\theta }{N}G^{\mu
_{1}\mu _{2}\ldots \mu _{N}}Z_{\mu _{1}}^{+}Z_{\mu _{2}}^{+}\ldots Z_{\mu
_{N}}^{+},  \label{generalized_special_duality}
\end{eqnarray}%
where as above
\begin{equation}
Z^{\pm }=x\pm \frac{1}{a}L,
\end{equation}%
and where $G^{\mu \mu _{1}\mu _{2}\ldots \mu _{N-1}}$ is general totally
symmetric traceless tensor. Again, it is an easy exercise to show that such
a transformation preserves the IHC  (see Appendix \ref{IHC2}, let
us note that for the invariance of IHC, $G^{\mu _{1}\mu _{2}\ldots \mu _{N}}$
need not to be traceless), i.e.%
\begin{equation}
\left[ \omega _{A}\right] _{\theta }=\omega .
\end{equation}%
Therefore (\ref{generalized_special_duality}) generates a consistent
transformation of the Galileon field. The obvious (though trivial; see
Appendix \ref{duality_on_special_galileon}) candidate for invariant Galileon
action can be constructed form the manifestly invariant $D+1$ form $\Omega
^{+}$defined in the previous section (cf (\ref{omega_plus_minus}))
\begin{equation}
\Omega ^{+}=\varepsilon _{\mu _{1}\mu _{2}\ldots \mu _{D}}\omega _{A}\wedge
\mathrm{d}Z^{+\mu _{1}}\wedge \mathrm{d}Z^{+\mu _{2}}\ldots \wedge \mathrm{d}%
Z^{+\mu _{D}}
\end{equation}%
and coincides with $S_{+}\left( a,c_{+}\right) $ introduced there. However,
the analogous form $\Omega ^{-}$ is not invariant under (\ref%
{generalized_special_duality}) and we cannot therefore combine it with $%
\Omega ^{+}$in order to get nontrivial action analogous to $S\left( \mathrm{i%
}a,-\mathrm{i}b\right) $.

Another less obvious invariant action can be constructed form the $D+1$ form%
\begin{eqnarray}
\Omega ^{L} &\equiv &\varepsilon _{\mu _{1}\mu _{2}\ldots \mu _{D-1}\nu
}\omega _{A}\wedge \mathrm{d}Z^{+\mu _{1}}\wedge \mathrm{d}Z^{+\mu
_{2}}\ldots \wedge \mathrm{d}Z^{+\mu _{D-1}}\wedge \mathrm{d}L^{\nu }  \notag
\\
&=&\left( -1\right) ^{D-1}\sum_{n=0}^{D-1}\left(
\begin{array}{c}
D-1 \\
n%
\end{array}%
\right) \frac{1}{a^{n}}\omega _{D+1}^{\left( n+2\right) }.
\end{eqnarray}%
The form $\Omega ^{L}$ transforms under (\ref{generalized_special_duality})
as%
\begin{eqnarray}
\left[ \Omega ^{L}\right] _{\theta } &=&\varepsilon _{\mu _{1}\mu _{2}\ldots
\mu _{D-1}\nu }\left[ \omega _{A}\right] _{\theta }\wedge \mathrm{d}%
Z_{\theta }^{+\mu _{1}}\wedge \mathrm{d}Z_{\theta }^{+\mu _{2}}\ldots \wedge
\mathrm{d}Z_{\theta }^{+\mu _{D-1}}\wedge \mathrm{d}L_{\theta }^{\nu }
\notag \\
&=&\varepsilon _{\mu _{1}\mu _{2}\ldots \mu _{D-1}\nu }\omega _{A}\wedge
\mathrm{d}Z^{+\mu _{1}}\wedge \mathrm{d}Z^{+\mu _{2}}\ldots \wedge \mathrm{d}%
Z^{+\mu _{D-1}}  \notag \\
&&\wedge \left( \mathrm{d}L^{\nu }-\theta \left( N-1\right) G\left(
Z^{+}\right) _{~\alpha }^{\nu }\mathrm{d}Z^{+\alpha }\right)
\end{eqnarray}%
where we have used%
\begin{equation}
L_{\theta }^{\nu }=L^{\nu }-\theta G^{\mu \mu _{1}\mu _{2}\ldots \mu
_{N-1}}Z_{\mu _{1}}^{+}Z_{\mu _{2}}^{+}\ldots Z_{\mu _{N-1}}^{+},
\end{equation}%
and where we abreviated%
\begin{equation}
G\left( Z^{+}\right) ^{\mu \nu }=G^{\mu \nu \alpha _{1}\alpha _{2}\ldots
\alpha _{N-2}}Z_{\alpha _{1}}^{+}Z_{\alpha _{2}}^{+}\ldots Z_{\alpha
_{N-2}}^{+}.
\end{equation}%
Note that%
\begin{equation}
\varepsilon _{\mu _{1}\mu _{2}\ldots \mu _{D-1}\nu }\mathrm{d}Z^{+\mu
_{1}}\wedge \mathrm{d}Z^{+\mu _{2}}\ldots \wedge \mathrm{d}Z^{+\mu
_{D-1}}\wedge \mathrm{d}Z^{\pm \alpha }=\frac{1}{D}\delta _{\nu }^{\alpha
}\varepsilon _{\mu _{1}\mu _{2}\ldots \mu _{D}}\mathrm{d}Z^{+\mu _{1}}\wedge
\mathrm{d}Z^{+\mu _{2}}\ldots \wedge \mathrm{d}Z^{+\mu }
\end{equation}%
and thus%
\begin{equation}
\left[ \Omega ^{L}\right] _{\theta }=\Omega ^{L}-\frac{\theta \left(
N-1\right) }{D}G\left( Z^{+}\right) _{~\nu }^{\nu }\varepsilon _{\mu _{1}\mu
_{2}\ldots \mu _{D}}\mathrm{d}Z^{+ \mu _{1}}\wedge \mathrm{d}Z^{+ \mu
_{2}}\ldots \wedge \mathrm{d}Z^{+ \mu_D }.
\end{equation}%
Therefore for traceless $G^{\mu _{1}\mu _{2}\ldots \mu _{N}}$ the form $%
\Omega ^{L}$ is invariant wirh respect to (\ref{generalized_special_duality}%
). Unfortunately, the action%
\begin{equation}
S^{L}=\int_{B_{D+1}}\Omega ^{L}|_{IHC}=\int \mathrm{d}^{D}x%
\sum_{n=2}^{D+1}d_{n}^{L}\left( a\right) \mathcal{L}_{n},  \label{S_L}
\end{equation}%
where%
\begin{equation}
d_{n}^{L}\left( a\right) =\frac{\left( -1\right) ^{D-1}}{n}\left(
\begin{array}{c}
D-1 \\
n-2%
\end{array}%
\right) \frac{1}{a^{n-2}},
\end{equation}%
can be shown to be dual to free theory \cite{Noller:2015rea}. Indeed, as we
show in Appendix \ref{duality_on_special_galileon}, under the duality
transformation (\ref{x_L_transformation}), (\ref{phi_transformation}) with
matrix $\boldsymbol{M}=\alpha _{D}\left( \sigma \right) $ (see (\ref%
{alpha_D_matrix})), we transform $d_{n}^{L}\left( a\right) $ into

\begin{equation}
d_{n}^{L}\left( a\right) _{\sigma }=\frac{\left( -1\right) ^{D-1}}{D}\frac{1%
}{n}\left(
\begin{array}{c}
D \\
n-1%
\end{array}%
\right) \left( \frac{1}{a}-2\theta \right) ^{n-2}.
\end{equation}%
Moreover, this transformation with $\sigma =1/2a$ (converting $S^{L}$ to the
free theory) transforms at the same time the action $S_{+}(a,c_{+})$ into the
trivial tadpole action $S_{1}(a,c_{+})$ (see (\ref{d_pm_sigma}) ). The
physical content of any linear combinations of the invariant actions $%
S_{+}(a,c_{+})$ and $S^{L}$ is therefore trivial.

Infinitesimal version of the transformation (\ref%
{generalized_special_duality}) reads after using the the inverse Higgs
constraint%
\begin{eqnarray}
x_{\theta }^{\mu } &=&x^{\mu }+\frac{\theta }{a}G^{\mu \mu _{1}\mu
_{2}\ldots \mu _{N-1}}\left( x_{\mu _{1}}+\frac{1}{a}\partial _{\mu
_{1}}\phi \left( x\right) \right) \ldots \left( x_{\mu _{N-1}}+\frac{1}{a}%
\partial _{\mu _{N-1}}\phi \left( x\right) \right)  \notag \\
\phi _{\theta }\left( x_{\theta }\right) &=&\phi \left( x\right) -\frac{%
\theta }{N}G^{\mu _{1}\mu _{2}\ldots \mu _{N}}\left( x_{\mu _{1}}+\frac{1}{a}%
\partial _{\mu _{1}}\phi \left( x\right) \right) \ldots \left( x_{\mu _{N}}+%
\frac{1}{a}\partial _{\mu _{N}}\phi \left( x\right) \right)  \notag \\
&&+\frac{\theta }{a}\partial _{\mu _{1}}\phi \left( x\right) G^{\mu _{1}\mu
_{2}\ldots \mu _{N}}\left( x_{\mu _{1}}+\frac{1}{a}\partial _{\mu _{1}}\phi
\left( x\right) \right) \ldots \left( x_{\mu _{N}}+\frac{1}{a}\partial _{\mu
_{N}}\phi \left( x\right) \right),
\end{eqnarray}%
or%
\begin{equation}
\phi _{\theta }\left( x\right) =\phi \left( x\right) -\frac{\theta }{N}%
G^{\mu _{1}\mu _{2}\ldots \mu _{N}}\left( x_{\mu _{1}}+\frac{1}{a}\partial
_{\mu _{1}}\phi \left( x\right) \right) \ldots \left( x_{\mu _{N}}+\frac{1}{a%
}\partial _{\mu _{N}}\phi \left( x\right) \right) .
\end{equation}%
In the latter form we recognize the \textquotedblleft extended Galileon
symmetry\textquotedblright\ of Noller, Sivanesan and von Strauss discussed
in \cite{Noller:2015rea} which (as disussed there and as we have shown here)
is in fact dual to the traceless polynomial shift symmetry of the free
theory with tadpole.

\section{Geometrical origin of the Special Galileon}

In this Section we give an alternative geometrical treatment of the
symmetries and dualities discussed above. We will show that, using the
complex coordinates (\ref{ZZbar}) introduced in the previous section, it is
possible, at least formally, to construct the Galileon Lagrangian in a way
analogous to the DBI one using the probe brane in higher dimensional space.
Within this approach we can interpret the Galileon field as a scalar degree
of freedom describing fluctuations of appropriately chosen $D-$dimensional
brane in $2D$ dimensional space. The latter has a pseudo-riemannian metric
with signature $(2,2D-2)$ and can be formally treated as a complexified
Minkowski space. Within such framework, the Galileon linear shift symmetry (%
\ref{linear_shift}) as well as the symmetry (\ref{hidden symmetry}) of the
Special Galileon originate from the symmetries of the target space and their
non-linear character is a consequence of gauge fixing of the
reparametrization freedom describing the embedding of the brane. This
construction can be easily repeated for the real coordinates (\ref{Zpm}) and
the transformations (\ref{duality_imaginary}) by means of appropriate
analytic continuation. Such an approach to Special Galileon, though purely
formal, will be useful for construction of the higher order Lagrangians
which are necessary as countertems when the theory is treated on the quantum
level.

Let us first describe the target space. Assume a $D-$dimensional complex
space $M_{\mathbf{\ \mathbb{C} }}^{D}=\mathbf{\ \mathbb{C} }^{D}$ with
coordinates
\begin{equation}
Z=X+\frac{\mathrm{i}}{\alpha }L,~~~~~\overline{Z}=X-\frac{\mathrm{i}}{\alpha
}L,
\end{equation}%
where $X^{\mu }$ and $L^{\mu }$ are real coordinates and $\alpha $ is fixed
real parameter (cf. (\ref{ZZbar})). Let us equip $M_{\mathbf{\ \mathbb{C} }%
}^{D}$ and with a hermitean form $h$ defined as%
\begin{equation}
h=\eta _{\mu \nu }\mathrm{d}Z^{\mu }\otimes \mathrm{d}\overline{Z}^{\mu }.
\end{equation}%
The real part of this form defines a metric with signature $\left(
2,2(D-1)\right) $ on $M_{\mathbf{C}}^{D}$ treated as a real $2D$ dimensional
space%
\begin{equation}
\mathrm{d}s^{2}=\mathrm{d}Z\cdot \mathrm{d}\overline{Z}=\eta _{\mu \nu
}\left( \mathrm{d}X^{\mu }\mathrm{d}X^{\nu }+\frac{1}{\alpha ^{2}}\mathrm{d}%
L^{\mu }\mathrm{d}L^{\nu }\right) ,  \label{target_metric}
\end{equation}%
while the imaginary part of $h$ \ generates a symplectic K\"{a}hler form
\begin{equation}
\omega =\frac{\mathrm{i}}{2}\eta _{\mu \nu }\mathrm{d}Z^{\mu }\wedge \mathrm{%
d}\overline{Z}^{\mu }=\frac{1}{\alpha }\eta _{\mu \nu }\mathrm{d}X^{\mu
}\wedge \mathrm{d}L^{\nu }.  \label{Kaehler form}
\end{equation}%
All the above forms are invariant with respect to the transformations%
\footnote{%
Transformation of $\overline{Z}$ is defined by means of complex conjugation.}%
\begin{equation}
Z^{\prime \mu }=R_{~\nu }^{\mu }Z^{\nu }+A^{\mu } ,  \label{complex_Poincare}
\end{equation}%
where the rotation matrix $R_{~\nu }^{\mu }$ $\in U(1,D-1)$, i.e. it
satisfies%
\begin{equation}
R^{+}\cdot \eta \cdot R=\eta ,
\end{equation}%
and the complex vector $A=c+\frac{\mathrm{i}}{\alpha }b\in \mathbf{\ \mathbb{%
C} }^{D}$ corresponds to a translation in $M_{\mathbf{\ \mathbb{C} }}^{D}$.
These transformations form a group $\mathbb{C} ^{D}\rtimes U(1,D-1)$ which
can be understood as a complex generalization of the Poincar\'{e} group $%
ISO(1,D-1)= \mathbb{R} ^{D}\rtimes O(1,D-1)$, which is its natural real
subgroup.

As an analoque of the DBI probe brane, we assume a $D-$dimensional real
Minkowski manifold $M_{\mathbf{\ \mathbb{R} }}^{D}$ embedded into $M_{%
\mathbf{\ \mathbb{C} }}^{D}$. The embedding is parametrized by real
parameters $\xi ^{\mu }$, $\mu =0,\ldots D-1$%
\begin{equation}
Z^{\mu }=Z^{\mu }\left( \xi \right) =X^{\mu }\left( \xi \right) +\frac{%
\mathrm{i}}{\alpha }L^{\mu }\left( \xi \right)
\end{equation}%
and we put additional constraint on the functions $Z^{\mu }\left( \xi
\right) $, namely we require that the K\"{a}hler form vanishes\footnote{%
Such submanifolds are known as the Lagrangian submanifolds.} on the brane%
\footnote{%
Here and in what follows we borrow the terminology from the DBI case.} $M_{
\mathbf{\ \mathbb{R} }}^{D}\ $%
\begin{equation}
\omega |_{M_{\mathbf{\ \mathbb{R} }}^{D}}=0.
\end{equation}%
Explicitly we get%
\begin{equation}
\frac{\partial Z}{\partial \xi ^{\mu }}\cdot \frac{\partial \overline{Z}}{%
\partial \xi ^{\nu }}=\frac{\partial \overline{Z}}{\partial \xi ^{\mu }}%
\cdot \frac{\partial Z}{\partial \xi ^{\nu }}  \label{constraint_omega}
\end{equation}%
or using the real coordinates%
\begin{equation}
\frac{\partial X}{\partial \xi ^{\mu }}\cdot \frac{\partial L}{\partial \xi
^{\nu }}=\frac{\partial L}{\partial \xi ^{\mu }}\cdot \frac{\partial X}{%
\partial \xi ^{\nu }}  \label{constraint_omega_1}
\end{equation}%
Let us note, that this constraint is invariant with respect to the
transformations (\ref{complex_Poincare}). On the brane $M_{\mathbf{\ \mathbb{%
R} }}^{D}$ we get a real induced metric%
\begin{equation}
\mathrm{d}s^{2}=\eta _{\alpha \beta }\frac{\partial Z^{\alpha }}{\partial
\xi ^{\mu }}\frac{\partial \overline{Z}^{\beta }}{\partial \xi ^{\nu }}%
\mathrm{d}\xi ^{\mu }\mathrm{d}\xi ^{\nu }\equiv g_{\mu \nu }\mathrm{d}\xi
^{\mu }\mathrm{d}\xi ^{\nu }.  \label{induced_metric}
\end{equation}%
Because both the target space metric (\ref{target_metric}) and the
constraint (\ref{constraint_omega}) are invariant with respect to (\ref%
{complex_Poincare}), the geometrical reparametrization invariant actions
built from the induced metric on $M_{\mathbf{\ \mathbb{R} }}^{D}$ , the
corresponding covariant derivatives and the intrinsic and external
curvatures will share the invariance (\ref{complex_Poincare}) too (see \cite%
{Hinterbichler:2010xn,Goon:2011qf} for the detailed description the brane
construction of effective actions and for discussion of their symmetry
properties).

Now we shall identify the Galileon as a scalar degree of freedom which
effectively describes the fluctuations of the brane $M_{\mathbf{\ \mathbb{C}
}}^{D}$ in the target space. We shall proceed in the way completely
analogous to the construction of the DBI-like actions. Let us first fix the
gauge freedom corresponding to the reparametrization invariance and
introduce new coordinates $x^{\mu }$ on the brane according to%
\begin{equation}
x^{\mu }=X^{\mu }\left( \xi \right) .  \label{new_parameters}
\end{equation}%
In such parametrization the embedding simplifies to%
\begin{equation}
X^{\mu }\left( x\right) =x^{\mu },\,\,\,L^{\mu }=L^{\mu }(x).
\label{gauge_condition}
\end{equation}%
The fluctuations of the brane are then effectively described by fields $%
L^{\mu }\left( x\right) $ living on it. However, we have to impose the
additional constraint (\ref{constraint_omega}) which further reduces the
number of effective degrees of freedom. Using the parametrization (\ref%
{new_parameters}), the constraint (\ref{constraint_omega_1}) simplifies to
the integrability condition for the field$\ L_{\mu }\left( x\right) =\eta
_{\mu \alpha }L^{\alpha }(x)$, namely
\begin{equation}
\partial _{\nu }L_{\mu }=\partial _{\mu }L_{\nu },  \label{integrability}
\end{equation}%
(here and in what follows, $\partial _{\mu }=\partial /\partial x^{\mu }$)
and as a consequence there exists $\phi(x)$ such that
\begin{equation}
L_{\mu }\left( x\right) =\partial _{\mu }\phi \left( x\right) .
\end{equation}%
In this gauge we have therefore the embeding of the brane described as%
\begin{equation}
X^{\mu }\left( x\right) =x^{\mu },\,\,\,L^{\mu }\left( x\right) =\eta ^{\mu
\nu }\partial _{\nu }\phi \left( x\right) .  \label{gauge_fixing}
\end{equation}%
Similarly to the case of DBI, we are left with one degree of freedom, which
is nothing else but the Galileon.\ To show this, let us discuss the
transformation properties of $\phi \left( x\right) $ with respect to the
transformations (\ref{complex_Poincare}). Let us note, that the gauge
condition (\ref{gauge_condition}) is not invariant under (\ref%
{complex_Poincare}). As a consequence, it is necessary to combine the target
space transformation with additional gauge transformation in order to ensure
the gauge fixing of the form of (\ref{gauge_fixing}) also for the
transformed brane. As a result, the field $\phi \left( x\right) $ will
generally transform non-linearly under (\ref{complex_Poincare}).

Let us first note that there is a residual symmetry%
\begin{equation}
\phi ^{\prime }\left( x\right) =\phi \left( x\right) +a,
\label{residual_shift}
\end{equation}%
which leaves the brane parametrization (\ref{gauge_fixing}) invariant. The
configurations of the brane are therefore parametrized rather by means of
the equivalence classes $\left[ \phi \right] =\left\{ \phi +a,a\in \mathbb{R}
\right\} $ modulo this symmetry. This means that the realization of the
transformations (\ref{complex_Poincare}) on the field $\phi \left( x\right) $
described in what follows are unique only up to the constant shift of the
field.

Under the complex translation the brane coordinates transform according to
\begin{equation}
Z^{\prime \mu }=Z^{\mu }+\left( c^{\mu }+\frac{\mathrm{i}}{\alpha }b^{\mu
}\right) ,
\end{equation}%
and thus the parameters $x^{\mu }$ shift by $a^{\mu }$ while the fields $%
L^{\mu }\left( x\right) $ shift by $b^{\mu }$, explicitly
\begin{equation}
X^{\prime \mu }\left( x\right) =x^{\mu }+c^{\mu },~~~L^{^{\prime }\mu
}\left( x\right) =\eta ^{\mu \nu }\partial _{\nu }\phi \left( x\right)
+b^{\mu }=\eta ^{\mu \nu }\partial _{\nu }\left[ \phi \left( x\right)
+b\cdot x\right] .
\end{equation}%
Let us define new parameters for the shifted brane
\begin{equation}
x^{\prime \mu }=x^{\mu }+c^{\mu },  \label{translation_reparametrization}
\end{equation}%
and define transformed field $\phi ^{\prime }\left( x^{\prime }\right) $ as
\begin{equation}
\phi ^{\prime }\left( x^{\prime }\right) =\phi \left( x\right) +b\cdot x.
\label{translation_phi_transformation}
\end{equation}%
Using $\partial _{\nu }^{\prime }=\partial _{\nu }$we get%
\begin{equation}
L^{^{\prime \prime }\mu }\left( x^{\prime }\right) \equiv L^{\prime \mu
}\left( x(x^{\prime })\right) =\eta ^{\mu \nu }\partial _{\nu }^{\prime
}\phi ^{\prime }\left( x^{\prime }\right) ,
\end{equation}%
and with the new field $\phi ^{\prime }\left( x^{\prime }\right) $ the gauge
(\ref{gauge_fixing}) is preserved. Therefore the complex translation is a
combination of the space-time translation and the Galileon transformation
corresponding to the linear shift of the Galileon field. The latter
corresponds to the purely imaginary shift.

Let us now write the general $U(1,D-1)$ rotation $R_{~\nu }^{\mu }$ in the
form%
\begin{equation}
R=\exp \left( \mathcal{M}+\mathrm{i}\mathcal{G}\right) =\Lambda +\mathrm{i}U,
\end{equation}%
where $\Lambda $ and $U$ are real matrices and $\mathcal{M}$ and $\mathcal{G}
$ are real generators. The latter satisfy%
\begin{eqnarray}
\eta _{\mu \rho }\mathcal{M}_{~\nu }^{\rho }+\eta _{\nu \rho }\mathcal{M}%
_{~\mu }^{\rho } &=&0 \\
\eta _{\mu \rho }\mathcal{G}_{~\nu }^{\rho }-\eta _{\nu \rho }\mathcal{G}%
_{~\mu }^{\rho } &=&0  \label{symmetry_condition}
\end{eqnarray}%
We get then for the transformed brane%
\begin{eqnarray}
X^{\prime \mu }\left( x\right) &=&\Lambda _{~\nu }^{\mu }x^{\nu }-\frac{1}{%
\alpha }U_{~\nu }^{\mu }L^{\nu }\left( x\right) =\Lambda _{~\nu }^{\mu
}x^{\nu }-\frac{1}{\alpha }U_{~\nu }^{\mu }\eta ^{\nu \rho }\partial _{\rho
}\phi \left( x\right) \\
L^{^{\prime }\mu }\left( x\right) &=&\frac{1}{\alpha }\Lambda _{~\nu }^{\mu
}L^{\nu }\left( x\right) +U_{~\nu }^{\mu }x^{\nu }=\frac{1}{\alpha }\Lambda
_{~\nu }^{\mu }\eta ^{\mu \nu }\partial _{\nu }\phi \left( x\right) +U_{~\nu
}^{\mu }x^{\nu }.
\end{eqnarray}%
For $U=0$ the matrices $\Lambda =\exp \mathcal{M}\in O(1,D-1)$ form a
subgroup of $U(1,D-1)$ and the above transformation reduces to the Lorentz
transformation with respect to which the field $\phi \left( x\right) $
transforms as a scalar. Indeed, after reparametrization%
\begin{equation}
x^{\prime \mu }=\Lambda _{~\nu }^{\mu }x^{\nu },
\end{equation}%
we immediately see that the new field $\phi ^{\prime }\left( x^{\prime
}\right) $ where
\begin{equation}
\phi ^{\prime }\left( x^{\prime }\right) =\phi \left( x\right) ,
\end{equation}%
preserves the gauge condition (\ref{gauge_fixing}). The remaining nontrivial
transformations $R=\exp \mathrm{i}{\mathcal{G}}$ are generated by matrices $%
\mathcal{G}_{~\nu }^{\rho }$ satisfying (\ref{symmetry_condition}), i.e.
they are of the form (\ref{U_matrix}). According to the results of the
previous section we can identify them with the hidden duality transformation
discussed there. The compensating gauge transformation corresponds to the
reparametrization%
\begin{equation}
x^{\prime \mu }=\Lambda _{~\nu }^{\mu }x^{\nu }-\frac{1}{\alpha }U_{~\nu
}^{\mu }L^{\nu }\left( x\right) =\Lambda _{~\nu }^{\mu }x^{\nu }-\frac{1}{%
\alpha }U_{~\nu }^{\mu }\eta ^{\nu \lambda }\partial _{\lambda }\phi \left(
x\right) ,  \label{rotation_reparametrization}
\end{equation}%
where $\Lambda =\mathrm{Re}\exp \mathrm{i}\mathcal{G}$ and $U=\mathrm{Im}%
\exp \mathrm{i}\mathcal{G}$ and the transformation of the field%
\begin{eqnarray}
\phi ^{\prime }\left( x^{\prime }\right) &=&\phi \left( x\right) -\frac{1}{2}%
x\cdot \partial \phi \left( x\right)  \notag \\
&&+\frac{1}{2}\eta _{\mu \nu }\left( \Lambda _{~\rho }^{\mu }x^{\rho }-\frac{%
1}{\alpha }U_{~\rho }^{\mu }\eta ^{\rho \alpha }\partial _{\alpha }\phi
\left( x\right) \right) \left( \frac{1}{\alpha }\Lambda _{~\sigma }^{\nu
}\eta ^{\sigma \beta }\partial _{\beta }\phi \left( x\right) +U_{~\sigma
}^{\nu }x^{\sigma }\right),  \label{rotation_phi_transformation}
\end{eqnarray}%
are exactly equivalent to (\ref{hidden symmetry}) by construction. Because
the latter transformation preserves the IHC, the gauge fixing condition (\ref%
{gauge_fixing}) is preserved as a consequence.

To conclude, the Galileon field can be formally understood as a scalar
degree of freedom describing fluctuations of the $D$ dimensional Lagrangian
brane in $2D-$dimensional real space $\mathbb{R} ^{2,2D-2}$ treated as a K%
\"{a}hler manifold. The symmetries and dualities of the special Galileon can
be explained as non-linear realization of the target space symmetry group $%
\mathbb{C} ^{D}\rtimes U(1,D-1)$. Its real subgroup $\mathbb{R}
^{1,D-1}\rtimes O(1,D-1)$ corresponds to the Poincar\'{e} symmetry $%
ISO(1,D-1)$ on the brane. Only $O(1,D-1)$ subgroup is realized linearly and
the linear shift transformations of the Galileon correspond to the pure
imaginary translation in the target space.

Note however, that within the target space symmetry group, the generators of
real and imaginary translation commute which is not the case of the
generators $P_{\mu }$ and $B_{\mu }$ of Poincar\'{e} translations and the
linear shift transformations of the Galileon filed respectively. Let us
remind (see (\ref{GAL})), that \ $[P_{\mu },B_{\mu }]=\mathrm{i}\eta _{\mu
\nu }A$ where $A$ is the generator of the residual constant shift symmetry
transformation $\phi \rightarrow \phi +a$ (see (\ref{residual_shift})).
However, as explained above, the brane configurations are classes of
equivalence modulo this constant shifts and on such classes the generator $A$
acts trivially, so that $P_{\mu }$ and $B_{\mu }$ effectively commute.

\section{Brane construction of the leading order Lagrangian}

Let us now show how to obtain the lowest order special Galileon action
within the above framework. It is natural to look for such an action in the
form\footnote{%
Writing the action as $S_{0}$ and denoting the coupling constant as $C_{0}$
we anticipate the more systematic notation for the higher order terms
explained in the next section.}
\begin{equation}
S_{0}=C_{0}\int_{M_{\mathbf{\ \mathbb{R} }}^{D}}\lambda +h.c.
\end{equation}
where $C\in \mathbb{C} $ is a complex normalization constant and $\lambda $
is appropriate $D-$form on the target space with as much symmetry as
possible with respect to the target space symmetry group. Of course, because
the basic Galileon Lagrangian is a Wess-Zumino term, we cannot expect that $%
\lambda $ will be completely $\mathbb{C} ^{D}\rtimes U(1,D-1)$ symmetric.

Natural building blocks for construction of such form $\lambda $ are the
holomorphic and antiholomorphic $D-$forms $\mathrm{d}^{D}Z$ and $\mathrm{d}%
^{D}\overline{Z}$ restricted to the brane, where (cf. (\ref{Omega}) and (\ref%
{Omega_bar}))
\begin{eqnarray}
\mathrm{d}^{D}Z &\equiv &\frac{1}{D!}\varepsilon _{\mu _{1}\mu _{2}\ldots
\mu _{D}}\mathrm{d}Z^{\mu _{1}}\wedge \mathrm{d}Z^{\mu _{2}}\ldots \wedge
\mathrm{d}Z^{\mu _{D}} \\
\mathrm{d}^{D}\overline{Z} &\equiv &\frac{1}{D!}\varepsilon _{\mu _{1}\mu
_{2}\ldots \mu _{D}}\mathrm{d}\overline{Z}^{\mu _{1}}\wedge \mathrm{d}%
\overline{Z}^{\mu _{2}}\ldots \wedge \mathrm{d}\overline{Z}^{\mu _{D}}.
\end{eqnarray}%
As we already know, these forms are invariant under $\mathbb{C} ^{D}\rtimes
SU(1,D-1)$ and under the general target space symmetry (\ref%
{complex_Poincare}) they take up a phase $\det R$ and $\det R^{+}$
respectively. However, because $\mathrm{d}^{D}Z$ and $\mathrm{d}^{D}%
\overline{Z}$ are exact, e.g.%
\begin{equation}
\mathrm{d}^{D}Z=\frac{1}{D}\mathrm{d}\left( \sigma \cdot Z\right) ,
\label{dZsigma}
\end{equation}%
where%
\begin{equation}
\sigma _{\mu }=\frac{1}{(D-1)!}\varepsilon _{\mu \mu _{1}\ldots \mu _{D-1}}%
\mathrm{d}Z^{\mu _{1}}\wedge \mathrm{d}Z^{\mu _{2}}\ldots \wedge \mathrm{d}%
Z^{\mu _{D-1}},  \label{sigma_form}
\end{equation}%
the simplest candidate of the invariant Lagrangian, namely $\mathcal{L}%
\mathrm{d}^{D}x=\mathrm{d}^{D}Z|_{M_{\mathbf{\ \mathbb{R} }}^{D}}$, is a
total derivative.

Another possible building block is the K\"{a}hler potential
\begin{equation}
\mathcal{K}=\frac{1}{2}Z\cdot \overline{Z}
\end{equation}%
which is invariant with respect to the homogenous subgroup $U(1,D-1)$ but
not with respect to the translations. Nevertheless let us construct the $%
SU(1,D-1)$ invariant $D-$form $\lambda $%
\begin{equation}
\lambda =2\mathcal{K}\mathrm{d}^{D}Z=\frac{1}{D!}\varepsilon _{\mu _{1}\mu
_{2}\ldots \mu _{D}}Z\cdot \overline{Z}\mathrm{d}Z^{\mu _{1}}\wedge \mathrm{d%
}Z^{\mu _{2}}\ldots \wedge \mathrm{d}Z^{\mu _{D}},
\end{equation}%
which transforms under translations $Z^{\prime \mu }=Z^{\mu }+$ $A^{\mu }$
according to%
\begin{equation}
\mathcal{\lambda }^{\prime }=Z^{\prime }\cdot \overline{Z}^{\prime }\mathrm{d%
}^{D}Z^{\prime }=\lambda +Z\cdot \overline{A}\mathrm{d}^{D}Z+A\cdot
\overline{Z}\mathrm{d}^{D}Z+A\cdot \overline{A}\mathrm{d}^{D}Z.
\end{equation}%
Thus the induced Lagrangian on the brane apparently breaks the Poincar\'{e}
translations as well as the linear Galileon shift. However, it can be easily
verified\footnote{%
Here we use formula (\ref{dZsigma}) and
\begin{eqnarray*}
\mathrm{d}Z^{\mu }\wedge \sigma _{\nu } &=&\delta _{\nu }^{\mu }\mathrm{d}%
^{D}Z. \\
0 &=&\eta _{\alpha \beta }\varepsilon _{\mu _{1}\ldots \mu _{D}}-\eta
_{\alpha \mu _{1}}\varepsilon _{\beta \mu _{2}\ldots \mu _{D}}+\ldots
+\left( -1\right) ^{D}\eta _{\alpha \mu _{D}}\varepsilon _{\beta \mu
_{1}\ldots \mu _{D-1}}
\end{eqnarray*}%
} that the following relation holds%
\begin{eqnarray}
&&A\cdot \overline{Z}\mathrm{d}^{D}Z+Z\cdot \overline{A}\mathrm{d}%
^{D}Z+A\cdot \overline{A}\mathrm{d}^{D}Z  \notag \\
&=&\mathrm{d}\left[ \frac{1}{D+1}Z\cdot \overline{A~}\sigma \cdot Z+\frac{1}{%
D-1}\left( A\cdot \overline{Z}~\sigma \cdot Z-Z\cdot \overline{Z~}\sigma
\cdot A\right) +\frac{1}{D}A\cdot \overline{A}~\sigma \cdot Z\right]  \notag
\\
&&+\frac{2\mathrm{i}}{D-1}\omega \wedge A\cdot \rho \cdot Z,
\end{eqnarray}%
where $\sigma _{\mu }$ is given by (\ref{sigma_form}), $\omega $ is the K%
\"{a}hler for (\ref{Kaehler form}) and we abbreviated%
\begin{equation}
\rho _{\mu \nu }=\frac{1}{(D-2)!}\varepsilon _{\mu \nu \mu _{1}\ldots \mu
_{D-2}}\mathrm{d}Z^{\mu _{1}}\wedge \mathrm{d}Z^{\mu _{2}}\ldots \wedge
\mathrm{d}Z^{\mu _{D-2}}.
\end{equation}%
Therefore $\lambda $ is invariant with respect to the complex translation
modulo exact form and an additional term proportional to the K\"{a}hler form
$\omega $. The latter vanishes by definition when restricted to the brane
and thus the integral of the form $\lambda $ over the brane $M_{\mathbf{\
\mathbb{R} }}^{D}$ is invariant with respect to the special Galileon
symmetry up to the phase $\det R$. The same is true for the complex
cojugated form $\overline{\lambda }=Z\cdot \overline{Z}\mathrm{d}^{D}%
\overline{Z}$. Let us therefore assume the following real action defined as%
\begin{equation}
S_{0}=\int_{M_{\mathbf{\ \mathbb{R} }}^{D}}\left[ C_{0}\lambda +C_{0}^{\ast }%
\overline{\lambda }\right] =\int \mathcal{L}_{0}\mathrm{d}^{D}x.
\end{equation}%
The individual terms of the corresponding Lagrangian $\mathcal{L}_{0}$ have
the right number of derivatives per field (namely $n_{\partial
}-n_{x}=2n_{\phi }-2$)\footnote{%
Having in mind integration by parts, we count here each explicit $x^{\mu }$
as \textquotedblleft inverse derivative\textquotedblleft $.$} as is required
\ for the basic Galileon Lagrangian and we therefore expect that $%
S_{C,C^{\ast }}$ can be identified with the special Galileon action
discussed in the previous section. Indeed, using the explicit expressions
for the form $\mathrm{d}^{D}Z|_{M_{\mathbf{\ \mathbb{R} }}^{D}}$ in terms of
the Galileon field%
\begin{equation}
\mathrm{d}^{D}Z|_{M_{\mathbf{\ \mathbb{R} }}^{D}}=\frac{1}{D!}\varepsilon
^{\mu _{1}\ldots \mu _{D}}\varepsilon ^{\nu _{1}\ldots \nu
_{D}}\sum_{n=0}^{D}\left(
\begin{array}{c}
D \\
n%
\end{array}%
\right) \left( \frac{\mathrm{i}}{\alpha }\right)
^{n}\prod\limits_{i=1}^{n}\partial _{\mu _{i}}\partial _{\nu _{i}}\phi
\prod\limits_{j=n+1}^{D}\eta _{\mu _{j}\nu _{j}}\mathrm{d}^{D}x,
\label{dZ_explicit}
\end{equation}%
performing integration by parts and setting
\begin{equation}
C_{0}=-\frac{\mathrm{i}}{4}\alpha ^{2}\left( D-1\right) !\mathrm{e}^{\mathrm{%
i}\beta }
\end{equation}%
we reproduce the action (\ref{S_real}).

\section{Higher order building blocks}

Once we have established the basic (lowest order) action, we can proceed
further and try to construct possible higher order Lagrangians. These are
necessary as countertems when the theory is treated on the quantum level. In
this section we will restrict ourselves to the case of the Special Galileon
$S(\alpha ,0)$ and $S(\mathrm{i}a,0)$ where $\alpha $ and $a$ are real
parameters. Apparently only such theories have\ a well defined quantum
version because for $\beta \neq 0$ or $b\neq 0$ the Lagrangians (\ref%
{action_alpha_xi}) and (\ref{Scpcm} ) contain nonzero tadpole term $\mathcal{%
L}_{1}$ which makes the perturbation theory ill-defined. In what follows we
take $\alpha $ real, the case $\alpha =\mathrm{i}a$ can be obtained in a
similar way (see Section \ref{section_analytic_continuation} for details).

The possible counterterms have to share the symmetry of the basic
Lagrangian, i.e. the symmetry with respect to the transformations (\ref%
{hidden symmetry}), (\ref{duality_imaginary}) with traceless $\mathcal{G}$.
In order to find such symmetric counterterms we need therefore to construct
their basic building blocks which are either invariant or have appropriate
covariant transformation properties under (\ref{hidden symmetry}) and (\ref%
{duality_imaginary}). In the previous section we have introduced the
geometrical interpretation of the special Galileon, which allows us to use
the well established machinery of the probe brane construction. Here this
approach gives us the invariants with respect to the special Galileon
symmetry.

The basic object of such a construction is the induced effective metric on
the brane
\begin{equation}
\mathrm{d}s^{2}=\mathrm{d}Z\cdot \mathrm{d}\overline{Z}|_{M_{\mathbf{\
\mathbb{R} }}^{D}}\equiv g_{\mu \nu }\mathrm{d}x^{\mu }\mathrm{d}x^{\nu }
\notag
\end{equation}%
where explicitly%
\begin{equation}
g_{\mu \nu }=\eta _{\mu \nu }+\frac{1}{\alpha ^{2}}\partial _{\mu }\partial
\phi \cdot \partial \partial _{\nu }\phi .
\end{equation}%
As a consequence of the invariance of $\mathrm{d}s^{2}=\mathrm{d}Z\cdot
\mathrm{d}\overline{Z}$ with respect to the target space symmetries, the
induced metric on the brane is invariant with respect to transformations (%
\ref{translation_reparametrization}), (\ref{translation_phi_transformation})
and (\ref{rotation_reparametrization}), (\ref{rotation_phi_transformation}).
Therefore the change of $g_{\mu \nu }$ under these symmetries reduces to a
covariant formula%
\begin{equation}
g_{\mu \nu }^{\prime }\left( x^{\prime }\right) =g_{\alpha \beta }\left(
x\right) \frac{\partial x^{\alpha }}{\partial x^{\prime \mu }}\frac{\partial
x^{\beta }}{\partial x^{\prime \nu }}.
\end{equation}%
Diffeomorphism invariants constructed from $g_{\mu \nu }$ are thus
automatically invariant with respect to the transformations (\ref{hidden
symmetry}), without any restrictions to $\mathcal{G}$ (i.e. $\mathcal{G}$
need not to be traceless). Let us now give a list of basic building blocks
for construction of such invariants.

The inverse metric $g^{\mu \nu }$ is represented as an infinite series
\footnote{%
In this and in the following formulas, the dot means contraction of the
Lorentz indices according to the flat metric $\eta _{\mu \nu }$, i.e. not
with $g_{\mu \nu }$. The symbol $\left( \partial \partial \phi \cdot
\partial \partial \phi \right) ^{\cdot n}$ denotes $n$-th matrix power of $%
\partial ^{\mu }\partial \phi \cdot \partial \partial _{\nu }\phi $. For
example, the $n=2$ term on the RHS of (\ref{inverse_g}) reads in detail
\begin{equation*}
-\frac{1}{\alpha ^{6}}\partial ^{\mu }\partial _{\mu _{1}}\phi \partial
^{\mu _{1}}\partial _{\mu _{2}}\phi \partial ^{\mu _{2}}\partial _{\mu
_{3}}\phi \partial ^{\mu _{3}}\partial ^{\nu }\phi
\end{equation*}%
where
\begin{equation*}
\partial ^{\alpha }\partial _{\beta }\phi =\eta ^{\alpha \sigma }\partial
_{\sigma }\partial _{\beta }\phi .
\end{equation*}%
}
\begin{equation}
g^{\mu \nu }=\eta ^{\mu \nu }+\sum_{n=1}^{\infty }\frac{\left( -1\right) ^{n}%
}{\alpha ^{2n}}\partial ^{\mu }\partial \phi \cdot \left( \partial \partial
\phi \cdot \partial \partial \phi \right) ^{\cdot \left( n-1\right) }\cdot
\partial \partial ^{\nu }\phi .  \label{inverse_g}
\end{equation}%
It is an easy exercise to calculate the other related objects, namely the
Christoffel symbols of the second kind
\begin{equation}
\Gamma _{\rho \sigma \mu }=\frac{1}{2}\left( \partial _{\mu }g_{\rho \sigma
}+\partial _{\sigma }g_{\rho \mu }-\partial _{\rho }g_{\sigma \mu }\right) =%
\frac{1}{\alpha ^{2}}\partial _{\mu }\partial _{\sigma }\partial \phi \cdot
\partial \partial _{\rho }\phi,
\end{equation}%
and the Riemann tensor
\begin{eqnarray}
R_{\alpha \beta \mu \nu } &=&\frac{1}{2}\left( \partial _{\beta }\partial
_{\mu }g_{\alpha \nu }+\partial _{\alpha }\partial _{\nu }g_{\beta \mu
}-\partial _{\alpha }\partial _{\mu }g_{\beta \nu }-\partial _{\beta
}\partial _{\nu }g_{\alpha \mu }\right) +g^{\rho \sigma }\left( \Gamma
_{\rho \beta \mu }\Gamma _{\sigma \alpha \nu }-\Gamma _{\rho \beta \nu
}\Gamma _{\sigma \alpha \mu }\right)  \notag \\
&=&\frac{1}{\alpha ^{2}}g^{\rho \sigma }\left( \partial _{\beta }\partial
_{\nu }\partial _{\rho }\phi \partial _{\sigma }\partial _{\mu }\partial
_{\alpha }\phi -\partial _{\alpha }\partial _{\nu }\partial _{\rho }\phi
\partial _{\sigma }\partial _{\mu }\partial _{\beta }\phi \right),
\label{Riemann_R}
\end{eqnarray}%
where we have used (\ref{inverse_g}).

Other building blocks are the components of the extrinsic curvature $K_{\mu
\nu }^{a}$ and the twist connection $\beta _{\mu a}^{b}$. These structures
are defined with help of the basis of tangent vectors $e_{\nu }^{A}$ and
normal vectors $n_{a}^{A}$ to the brane (see \cite{Hinterbichler:2010xn} for
a general construction) as
\begin{eqnarray}
e_{\mu }^{B}\nabla _{B}e_{\nu }^{A} &=&\Gamma _{\mu \nu }^{\rho }e_{\rho
}^{A}-K_{\mu \nu }^{a}n_{a}^{A} \\
e_{\mu }^{B}\nabla _{B}n_{a}^{A} &=&\beta _{\mu a}^{b}n_{b}^{A}+K_{a\mu
}^{~~\nu }e_{\nu }^{A}.
\end{eqnarray}%
Here the capital latin letters $A\equiv \left( \alpha ,\beta \right) ,\ldots
$corespond to the target space indices with respect to the coordinates $%
Y^{A}\equiv \left( X^{\alpha },L^{\beta }\right) $, small latin letters $%
a,b,\ldots $ denote the $D-$bein index of the normal space to the brane, the
greek letters $\mu ,\nu ,\ldots $ refer to the coordinates $x^{\mu }$ on the
brane and\footnote{%
Note that the target space is flat.} $\nabla _{A}=\partial _{A}$. The
tangent vectors $e_{\mu }$ are given as\footnote{%
In a more detailed notation%
\begin{equation*}
e_{\mu }=\frac{\partial }{\partial X^{\mu }}+\eta ^{\beta \nu }\partial
_{\mu }\partial _{\nu }\phi \frac{\partial }{\partial L^{\beta }}.
\end{equation*}%
}%
\begin{equation}
e_{\mu }^{A}=\frac{\partial Y^{A}}{\partial x^{\mu }}=\left( \delta _{\mu
}^{\alpha },\eta ^{\beta \nu }\partial _{\mu }\partial _{\nu }\phi \right) ,
\end{equation}%
while the normal vectors have to satisfy the orhogonality conditions%
\begin{equation}
G_{AB}n_{a}^{A}n_{b}^{B}=\eta _{ab},~~~G_{AB}n_{a}^{A}e_{\beta }^{B}=0,
\end{equation}%
where $G_{AB}=\mathrm{diag}\left( \eta _{\alpha \beta },\left( 1/\alpha
^{2}\right) \eta _{\delta \gamma }\right) $ is the target space metric in
the coordinates $Y^{A}\equiv \left( X^{\alpha },L^{\beta }\right) $. The $D-$%
bein of the normal vectors $n_{a}^{A}$ can be constructed using the $D-$%
dimensional vectors $m_{a}$ satisfying
\begin{equation}
g_{\alpha \beta }m_{a}^{\alpha }m_{b}^{\beta }=\eta _{ab},~~~\eta
^{ab}m_{a}^{\alpha }m_{b}^{\beta }=g^{\alpha \beta }.
\label{m_a_normalization}
\end{equation}%
In terms of these vectors we obtain%
\begin{equation}
n_{a}^{A}=\left( -\frac{1}{\alpha }\partial ^{\alpha }\partial \phi \cdot
m_{a},\alpha m_{a}^{\beta }\right) .
\end{equation}%
Note however, that the constraint (\ref{m_a_normalization}) does not fix the
vectors $m_{a}$ unambiguously. There is a freedom which allows us to
redefine the $D-$bein using a local Lorentz transformation acting on the
small Latin indices%
\begin{equation}
m_{a}^{\prime }=\Lambda _{a}^{~b}m_{b},~~~~~\Lambda _{a}^{~c}\eta
_{cd}\Lambda _{b}^{~d}=\eta _{ab}.  \label{gauge_transformation}
\end{equation}%
This give rise to additional gauge invariance of the construction we have to
take care of. The twist connection $\beta _{\mu a}^{b}$ takes a role of the
compensating gauge field corresponding to this gauge transformation.

Introducing a dual basis $e_{A}^{\mu }$ and $n_{A}^{a}$
\begin{equation}
e_{A}^{\mu }e_{\nu }^{A}=\delta _{\nu }^{\mu },~~~n_{A}^{a}n_{b}^{A}=\delta
_{b}^{a},~~~n_{A}^{a}e_{\mu }^{A}=n_{a}^{A}e_{A}^{\mu }=0,~~~e_{\mu
}^{A}e_{B}^{\mu }+n_{a}^{A}n_{B}^{a}=\delta _{B}^{A},
\end{equation}%
explicitly%
\begin{equation}
e_{A}^{\mu }=g^{\mu \sigma }\left( \eta _{\sigma \alpha },\frac{1}{\alpha
^{2}}\partial _{\sigma }\partial _{\beta }\phi \right) ,~~~~~~n_{A}^{a}=%
\frac{1}{\alpha }\eta ^{ab}\left( -m_{b}\cdot \partial \partial _{\alpha
}\phi ,\eta _{\beta \sigma }m_{b}^{\sigma }\right) ,
\end{equation}%
we can express $K_{\mu \nu }^{a}$ and $\beta _{\mu a}^{b}$ as follows%
\begin{eqnarray}
K_{\mu \nu }^{a} &=&-n_{A}^{a}e_{\mu }^{B}\nabla _{B}e_{\nu }^{A}=-\frac{1}{%
\alpha }\eta ^{ab}m_{b}\cdot \partial \partial _{\mu }\partial _{\nu }\phi \\
\beta _{\mu a}^{b} &=&n_{A}^{b}e_{\mu }^{B}\nabla _{B}n_{a}^{A}=\eta
^{bc}\left( m_{c}^{\alpha }g_{\alpha \beta }\partial _{\mu }m_{a}^{\beta }+%
\frac{1}{\alpha ^{2}}m_{c}\cdot \partial \partial \phi \cdot \partial
\partial _{\mu }\partial \phi \cdot m_{\alpha }\right) .
\end{eqnarray}%
and $K_{a\mu }^{~~\nu }=\eta _{ab}g^{\nu \sigma }K_{\mu \sigma }^{b}$. Under
the gauge transformation (\ref{gauge_transformation}) these objects
transform as (cf. (\ref{m_a_normalization}))
\begin{eqnarray}
K_{a\mu }^{~~\nu \prime } &=&\Lambda _{a}^{~b}K_{b\mu }^{~~\nu } \\
\beta _{\mu ab}^{\prime } &=&\eta _{ac}\beta _{\mu b}^{c\prime }=\Lambda
_{a}^{~c}\Lambda _{b}^{~d}\beta _{\mu cd}+\Lambda _{a}^{~c}\eta
_{cd}\partial _{\mu }\Lambda _{b}^{~d},
\end{eqnarray}%
while under the duality transformation (\ref{hidden symmetry}) they
transform covariantly as tensors with corresponding Greek indices.

However, not all these objects are in fact independent. Note that as a
consequence of (\ref{m_a_normalization}) we get independently on the choice
of $m_{a}$%
\begin{equation}
\eta _{ab}K_{\mu \alpha }^{a}K_{\nu \beta }^{b}=\frac{1}{\alpha ^{2}}%
\partial _{\mu }\partial _{\alpha }\partial _{\rho }\phi g^{\rho \sigma
}\partial _{\sigma }\partial _{\nu }\partial _{\beta }\phi,
\label{K_contracted}
\end{equation}%
and thus (see (\ref{Riemann_R})) we get the Gauss formula
\begin{equation}
R_{\alpha \beta \mu \nu }=\eta _{ab}\left( K_{\mu \alpha }^{a}K_{\nu \beta
}^{b}-K_{\mu \beta }^{a}K_{\nu \alpha }^{b}\right),  \label{gauss}
\end{equation}
which is valid in this form for the flat target space. Also, the curvature
of the twist connection%
\begin{equation}
\varphi _{\mu \nu b}^{a}=\partial _{\mu }\beta _{\nu b}^{a}-\partial _{\nu
}\beta _{\mu b}^{a}+\beta _{\mu c}^{a}\beta _{\nu b}^{c}-\beta _{\nu
c}^{a}\beta _{\mu b}^{c}
\end{equation}%
can be expressed in terms of the extrinsic curvature using the flat target
space form of the Ricci equation%
\begin{equation}
\varphi _{\mu \nu b}^{a}=g^{\rho \sigma }\eta _{bc}\left( K_{\mu \rho
}^{a}K_{\nu \sigma }^{c}-K_{\nu \rho }^{a}K_{\mu \sigma }^{c}\right) .
\label{codazzi}
\end{equation}%
Finally, we get also the Codazzi equation in the form%
\begin{equation}
D_{\mu }K_{\nu \alpha }^{a}-D_{\nu }K_{\mu \alpha }^{a}=0.
\end{equation}%
Here the covariant derivative $D_{\mu }$ acts also on the $D-$bein index as%
\begin{equation}
D_{\mu }K_{\nu \alpha }^{a}=\partial _{\mu }K_{\nu \alpha }^{a}-\Gamma _{\mu
\nu }^{\sigma }K_{\sigma \alpha }^{a}-\Gamma _{\mu \alpha }^{\sigma }K_{\nu
\sigma }^{a}+\beta _{\mu b}^{a}K_{\nu \alpha }^{b},
\end{equation}%
and analogically for other objects carrying the small Latin indices. We can
e.g. easily verify, that the vectors $m_{a}$ are covariantly constat%
\begin{equation}
D_{\mu }m_{a}^{\alpha }=\partial _{\mu }m_{a}^{\alpha }+\Gamma _{\mu \nu
}^{\alpha }m_{a}^{\nu }-\beta _{\mu a}^{b}m_{b}^{\alpha }=0.
\end{equation}

Note that, using an expansion%
\begin{equation}
\left( 1+x\right) ^{-1/2}=\sum_{n=0}^{\infty }\frac{\left( \frac{1}{2}%
-n\right) _{n}}{n!}x^{n},
\end{equation}%
where $\left( a\right) _{n}=\Gamma \left( a+n\right) /\Gamma \left( a\right)
$ is the Pochammer symbol, a particular solution of the constraints (\ref%
{m_a_normalization}) can be formally written in terms of an infinite series%
\begin{equation}
m_{a}^{\alpha }=\eta ^{\alpha \beta }\left( \eta _{a\beta
}+\sum_{n=1}^{\infty }\frac{\left( \frac{1}{2}-n\right) _{n}}{n!}\frac{1}{%
\alpha ^{2n}}\left[ \left( \partial \partial \phi \right) ^{\cdot n}\right]
_{a\beta }\right) .  \label{m_a_solution}
\end{equation}%
This choice of $m_{a}$ corresponds to a particular gauge fixing of the local
$O(1,D-1)$ invariance (\ref{gauge_transformation}).

The last building blocks are the invariant measures on the brane. In the
previous section we have discussed the forms $\mathrm{d}^{D}Z$ and $\mathrm{d%
}^{D}\overline{Z}$ which are invariant under (\ref{hidden symmetry}) up to
the phase. From the induced metric $g_{\mu \nu }$ we can construct strictly
invariant volume element%
\begin{equation}
\mathrm{d}^{D}x\sqrt{\left\vert \det \left( g_{\mu \nu }\right) \right\vert }%
=\mathrm{d}^{D}x\left\vert \det \left( \frac{\partial Z^{\alpha }}{\partial
x^{\mu }}\right) \right\vert \equiv \sqrt{\mathrm{d}^{D}Z\mathrm{d}^{D}%
\overline{Z}}.
\end{equation}%
After some algebra we get%
\begin{equation}
\sqrt{\mathrm{d}^{D}Z\mathrm{d}^{D}\overline{Z}}=\mathrm{d}%
^{D}x\sum_{M=0}^{\infty }\left( -\frac{1}{\alpha ^{2}}\right)
^{M}\sum_{N=0}^{\infty }\frac{1}{N!}\sum_{n_{i}\geq
1,~\sum_{j=1}^{N}n_{j}=M}\prod\limits_{k=1}^{N}\left( -\frac{1}{2n_{k}}\eta
^{\alpha \beta }\left[ \left( \partial \partial \phi \right) ^{\cdot 2n_{k}}%
\right] _{\alpha \beta }\right) .  \label{det_g}
\end{equation}%
This finishes our list of the basic building block of the higher order
Lagrangians.

\begin{figure}[t]
\begin{center}
\epsfig{width=0.9\textwidth,figure=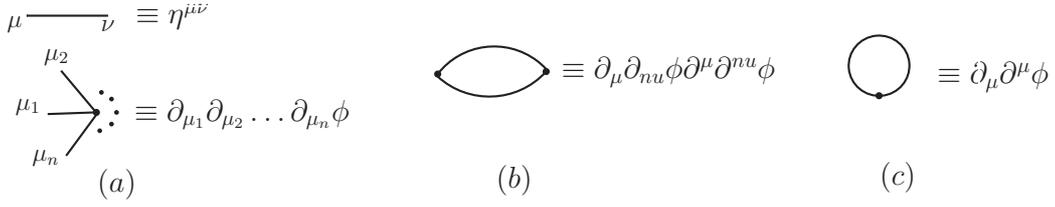}
\end{center}
\caption{The rules for graphical description of the invariant Lagrangians.
The basic graphical building block correspond to (a) while simple examples
are depicted as (b) and (c).}
\label{figure_graphic_language}
\end{figure}

\begin{figure}[tbp]
\begin{center}
\epsfig{width=0.9\textwidth,figure=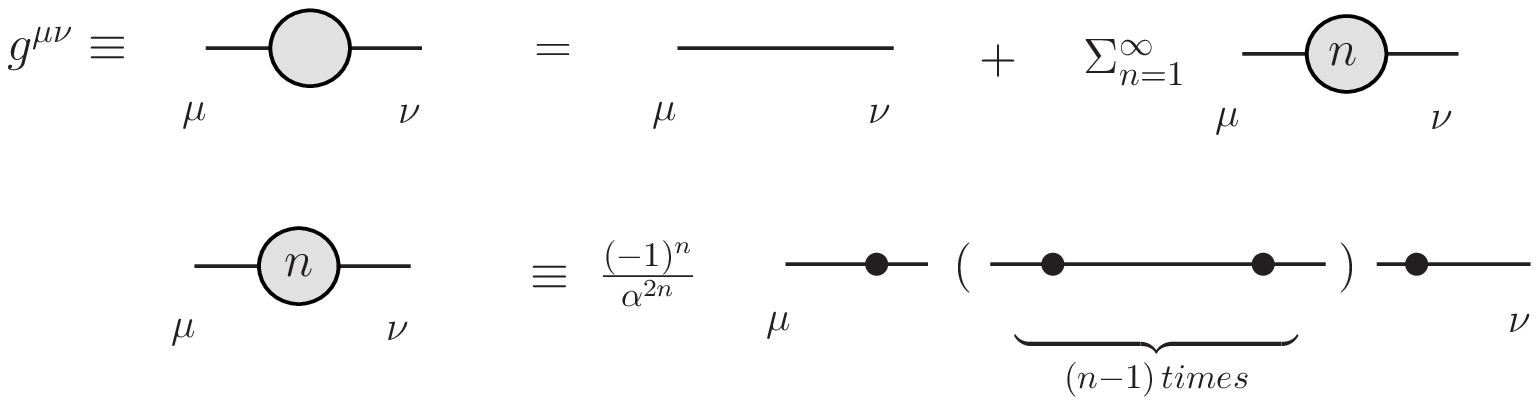} \vspace*{-0.5cm}
\end{center}
\caption{The graphical representation of the inverse effective metric $g^{%
\protect\mu \protect\nu }$. }
\label{figure_inverse_g}
\end{figure}

\begin{figure}[t]
\begin{center}
\epsfig{width=0.75\textwidth,figure=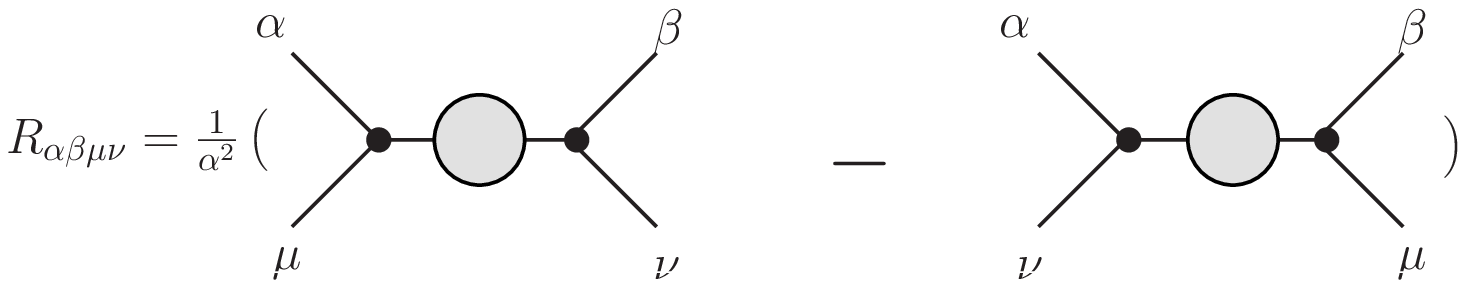} \vspace*{-0.5cm}
\end{center}
\caption{The graphical representation of the Riemann tensor $R_{\protect\mu
\protect\nu \protect\alpha \protect\beta }$. The blob represents the inverse
metric $g^{\protect\rho \protect\sigma }$ (cf. Fig. \protect\ref%
{figure_inverse_g}).}
\label{figure_Riemann_R}
\end{figure}

All these object can be vizualized and easily manipulated using an efficient
graphical representation developed in \cite{Griffin:2014bta}. Each field $%
\phi $ is represented with a point and the derivative $\partial _{\mu }$
acting on $\phi $ is depicted as a line starting at the point representing $%
\phi $ and carrying a corresponding Lorentz index $\mu $. The flat metric $%
\eta ^{\mu \nu }$ is drawn as line with indices $\mu $, $\nu $ and
contraction of the Lorentz indices is then represented as an internal line
connecting the points adjacent to the contracted derivatives. Also
integration by parts can be visualized within the graphical language: one
simply disconnect one end of a chosen line from the corresponding point and
the free end of this line attaches successively to all other points in the
graph. The sum of resulting graphs is then taken with additional minus sign.
The basic graphical rules as well as simple examples of their application
are shown in figure \ref{figure_graphic_language}. The graphical
representation of inverse metric and the Riemann tensor are then depicted in
figures \ref{figure_inverse_g} and \ref{figure_Riemann_R}.
\begin{figure}[t]
\begin{center}
\epsfig{width=0.75\textwidth,figure=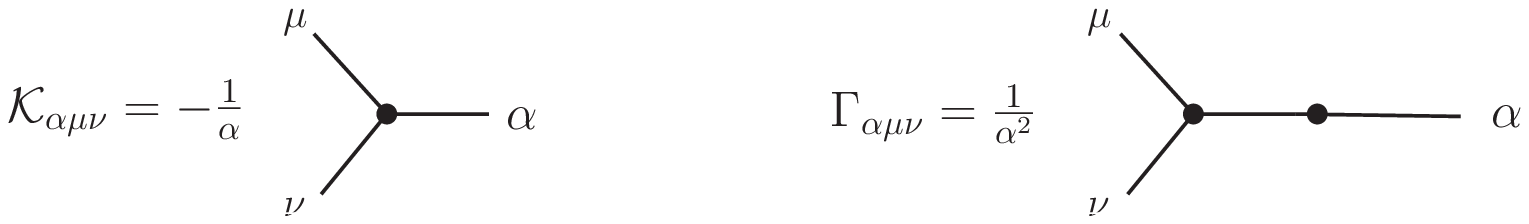} \vspace*{-0.5cm}
\end{center}
\caption{The graphical representation of the extrinsic curvature ${\mathcal{K%
}}_{\protect\alpha \protect\mu \protect\nu }$ and the Christoffel symbol $%
\Gamma_{\protect\alpha \protect\mu \protect\nu }$ .}
\label{figure_d_Z}
\end{figure}
\begin{figure}[t]
\begin{center}
\epsfig{width=0.9\textwidth,figure=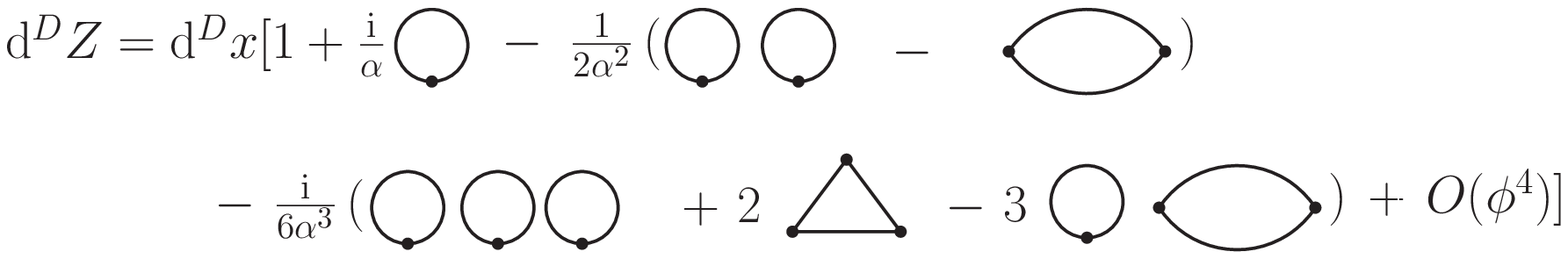}
\end{center}
\caption{The measure ${\mathrm{d}}^DZ$ in terms of the Galileon field.}
\label{figure_measure_Z}
\end{figure}
\begin{figure}[t]
\begin{center}
\epsfig{width=0.9\textwidth,figure=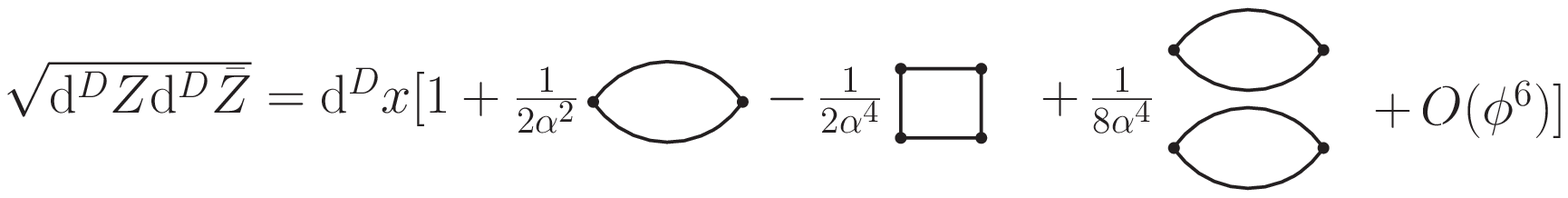}
\end{center}
\caption{The measure $\protect\sqrt{{\mathrm{d}}^DZ{\mathrm{d}}^D\bar{Z}}$
in terms of the Galileon field.}
\label{figure_measure_Z_barZ}
\end{figure}

\section{Higher order Lagrangians}

Having established the basic building blocks, let us proceed to the
construction of the higher order Lagrangians. According to the general
prescription for the probe brane action \cite%
{Hinterbichler:2010xn,Goon:2011qf}, the most general case can be written in
the form
\begin{equation}
S_{inv}=\int_{M_{\mathbf{\ \mathbb{R}}}^{D}}\mathrm{d}^{D}Z\mathcal{L}%
_{Z}+\int_{M_{\mathbf{\ \mathbb{R}}}^{D}}\mathrm{d}^{D}\overline{Z}\mathcal{L%
}_{\overline{Z}}+\int_{M_{\mathbf{\ \mathbb{R}}}^{D}}\sqrt{\mathrm{d}^{D}Z%
\mathrm{d}^{D}\overline{Z}}\mathcal{L}_{Z\overline{Z}}.
\end{equation}%
Here the functions $\mathcal{L}_{Z}$, $\mathcal{L}_{\overline{Z}}\equiv
\mathcal{L}_{Z}^{\ast }$ and $\mathcal{L}_{Z\overline{Z}}=\mathcal{L}_{Z%
\overline{Z}}^{\ast }$ are difeomorphism invariants and invariants with
respect to the local gauge transformation (\ref{gauge_transformation})
constructed from the building blocks listed in the previous section and
their covariant derivatives. The action $S_{inv}$ is then invariant with
respect to the transformations (\ref{hidden symmetry}) with traceless $%
G^{\mu \nu }$ and form-invariant (i.e. invariant up to a change of
couplings) for general $G^{\mu \nu }$.

Thanks to the Gauss-Codazzi formulae (\ref{gauss}) and (\ref{codazzi}), it
is sufficient to use only the extrinsic curvature $K_{\mu \nu }^{a}$, its
covariant derivatives and the $D-$bein $m_{a}^{\alpha }$. The invariants are
then obtained contracting the Greek indices of such building blocks with
respect to the induced metric $g_{\mu \nu }$ and its inverse $g^{\mu \nu }$,
while the small Latin indices have to be contracted using the flat metric $%
\eta _{ab}$. Because $m_{a}$ is the $D-$bein with respect to the induced
metric $g_{\mu \nu }$, we can convert freely the Greek indices into small
Latin indices and vice versa. For instance, instead of $K_{\mu \nu }^{a}$ we
can use solely the following rank-three tensor $\mathcal{K}_{\alpha \mu \nu
} $
\begin{equation}
\mathcal{K}_{\alpha \mu \nu }=g_{\alpha \beta }m_{a}^{\beta }K_{\mu \nu
}^{a}=-\frac{1}{\alpha }\partial _{\alpha }\partial _{\mu }\partial _{\nu
}\phi ,
\end{equation}%
and its covariant derivatives as the basic building blocks. In terms of $%
\mathcal{K}_{\alpha \mu \nu }$ the Gauss formula reads (see figures \ref%
{figure_d_Z} and \ref{figure_Riemann_R})%
\begin{equation}
R_{\alpha \beta \mu \nu }=g^{\rho \sigma }\left( \mathcal{K}_{\rho \mu
\alpha }\mathcal{K}_{\sigma \nu \beta }-\mathcal{K}_{\rho \mu \beta }%
\mathcal{K}_{\sigma \nu \alpha }\right) .  \label{gauss_1}
\end{equation}

\subsection{Hierarchy of the countertems}
Before giving some explicit examples of the higher order actions, let us
briefly comment on the hierarchy of the counterterms. As explained in \cite%
{Kampf:2014rka}, in the Galileon theories the higher order Lagrangians with $%
n_{\partial }$ derivatives and $n_{\phi }$ fields can be classified
according to the index $\delta $ where%
\begin{equation}
\delta =n_{\partial }-2n_{\phi }+2.
\end{equation}%
In what follows we will denote therefore the higher order actions and
Lagrangians as $S_{\delta }$ and $\mathcal{L}_{\delta }$ respectively in
order to indicate the index of their vertices. While all the terms of the
basic Lagrangian have $\delta =0$, the indices of the countertems which are
necessary to cancel UV divergences of the graph with $L$ loops and vertices $%
V_{i}$ with indices $\delta _{i}$ in $D$ spacetime dimensions are \cite%
{Kampf:2014rka}
\begin{equation}
\delta _{CT}=\left( D+2\right) L+\sum_{i}\delta _{i}.  \label{power_counting}
\end{equation}%
Therefore the first Lagrangians which are renormalized by loop corrections
has index $\delta =D+2$.

The assignment of the index $\delta $ to a concrete
term in the Lagrangian is easy. Note that $\sqrt{\mathrm{d}^{D}Z\mathrm{d}%
^{D}\overline{Z}}$, $\mathrm{d}^{D}Z$ and $\mathrm{d}^{D}\overline{Z}$
depend only on second derivatives of $\phi $ and therefore they do not
contribute to $\delta $ at all. The same is true for $g^{\mu \nu }$, $g_{\mu
\nu }$, while each $\mathcal{K}_{\alpha \mu \nu }$ as well as each covariant
derivative $D_{\mu }$ increases $\delta $ by one. As a result
\begin{equation}
\delta =n_{\mathcal{K}}+n_{D}+2  \label{index}
\end{equation}%
where $n_{\mathcal{K}}$ is number of $\mathcal{K}_{\alpha \mu \nu }$ and $%
n_{D}$ is number of covariant derivatives. In what follows we will discuss
several lowest $\delta $ actions in more detail.

\subsection{Lagrangians with $\protect\delta =2$}

Let us start with the action $S_{2}$. Because, unlike $\mathrm{d}^{D}Z$ and $%
\mathrm{d}^{D}\overline{Z}$, the invariant volume element $\sqrt{\mathrm{d}%
^{D}Z\mathrm{d}^{D}\overline{Z}}$ is not trivial when integrated over the
brane, the next to lowest action with $\delta =2$, i.e. with $n_{\partial
}=2n_{\phi }$ derivatives, can be constructed as
\begin{equation}
S_{2}=B_{2}\int_{M_{\mathbf{\ \mathbb{R} }}^{D}}\sqrt{\mathrm{d}^{D}Z\mathrm{%
d}^{D}\overline{Z}}.
\end{equation}%
Such a term is unique up to a real constant $B_{2}$ and corresponds to the
cosmological constant term for the induced metric $g_{\mu \nu }$. Using (\ref%
{det_g}) we get explicitly%
\begin{eqnarray}
\mathcal{L}_{2} &=&B_{2}\sqrt{\left\vert \det \left( g_{\mu \nu }\right)
\right\vert }=B_{2}\left[ 1+\frac{1}{2\alpha ^{2}}\left\langle \partial
\partial \phi \cdot \partial \partial \phi \right\rangle \right.  \notag \\
&&\left. -\frac{1}{2\alpha ^{4}}\left\langle \partial \partial \phi \cdot
\partial \partial \phi \cdot \partial \partial \phi \cdot \partial \partial
\phi \right\rangle +\frac{1}{8\alpha ^{4}}\left\langle \partial \partial
\phi \cdot \partial \partial \phi \right\rangle ^{2}+O\left( \phi
^{6}\right) \right] .  \label{det_g_explicit}
\end{eqnarray}%
Here and in what follows we abreviated by $\left\langle .\right\rangle $ a
trace of rank two Minkowski tensor with respect to the flat metric $\eta
^{\mu \nu }$. The action $S_{2}$ introduces a higher order kinetic term as
well as an infinite tower of related interaction terms. The constant $B_{2}$
is not renormalized by loop corrections in any dimensions.

\begin{figure}[t]
\begin{center}
\epsfig{width=0.9\textwidth,figure=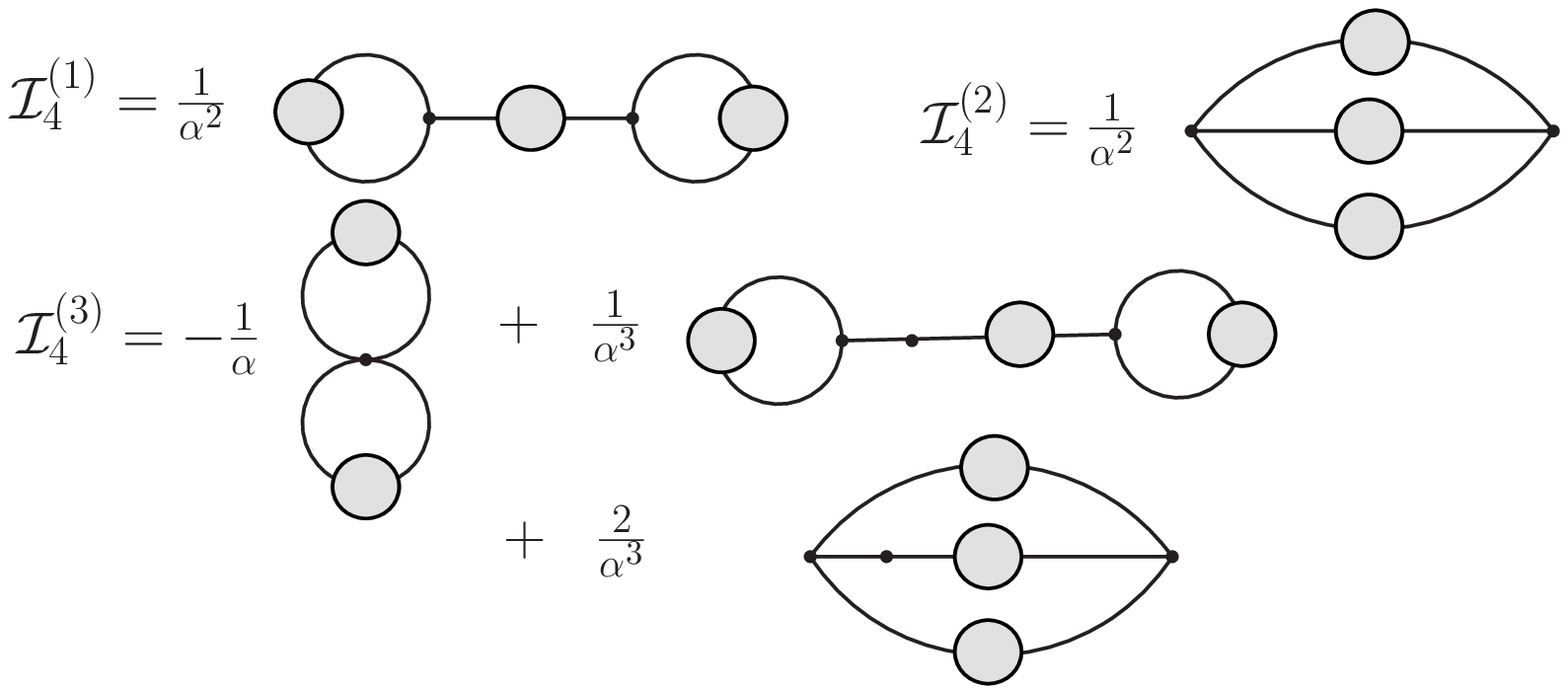}
\end{center}
\caption{The three independent invariants for construction of the general $%
\protect\delta=4$ Lagrangian. The blob corresponds to the inverse metric
tensor (see Fig.\protect\ref{figure_inverse_g})}
\label{figure_delta4_invariants}
\end{figure}

\subsection{Lagrangians with $\protect\delta =4$}

Next invariant action correspond to $\delta =4$, i.e. $n_{\partial
}=2n_{\phi }+2$. There are three idependent invariants $\mathcal{I}%
_{4}^{\left( j\right) }$ $j=1,2,3$ with $n_{\mathcal{K}}+n_{D}=2$ available
for the construction of $\mathcal{L}_{Z}$, $\mathcal{L}_{\overline{Z}}$ and $%
\mathcal{L}_{Z\overline{Z}}$, namely (see Fig.\ref{figure_delta4_invariants}
for graphical representation)%
\begin{eqnarray}  \label{I4(3)}
\mathcal{I}_{4}^{\left( 1\right) } &=&g^{\alpha \beta }g^{\mu \nu }g^{\rho
\sigma }\mathcal{K}_{\alpha \mu \nu }\mathcal{K}_{\beta \rho \sigma }=\frac{1%
}{\alpha ^{2}}g^{\alpha \beta }g^{\mu \nu }g^{\rho \sigma }\partial _{\alpha
}\partial _{\mu }\partial _{\nu }\phi \partial _{\beta }\partial _{\rho
}\partial _{\sigma }\phi  \label{I4(1)} \\
\mathcal{I}_{4}^{\left( 2\right) } &=&g^{\alpha \beta }g^{\mu \rho }g^{\nu
\sigma }\mathcal{K}_{\alpha \mu \nu }\mathcal{K}_{\beta \rho \sigma }=\frac{1%
}{\alpha ^{2}}g^{\alpha \beta }g^{\mu \rho }g^{\nu \sigma }\partial _{\alpha
}\partial _{\mu }\partial _{\nu }\phi \partial _{\beta }\partial _{\rho
}\partial _{\sigma }\phi  \label{I4(2)} \\
\mathcal{I}_{4}^{\left( 3\right) } &=&g^{\alpha \beta }g^{\mu \nu }D_{\alpha
}\mathcal{K}_{\beta \mu \nu }=\frac{1}{\sqrt{\left\vert g\right\vert }}%
\partial _{\alpha }\left( \left\vert g\right\vert g^{\alpha \beta }\mathcal{K%
}_{\beta \mu \nu }g^{\mu \nu }\right)  \notag \\
&=&-\frac{1}{\alpha }g^{\alpha \beta }g^{\mu \nu }\partial _{\alpha
}\partial _{\beta }\partial _{\mu }\partial _{\nu }\phi +\frac{1}{\alpha ^{3}%
}g^{\rho \sigma }\left( g^{\alpha \beta }g^{\mu \nu }+2g^{\alpha \mu
}g^{\beta \nu }\right) \partial _{\alpha }\partial _{\beta }\partial \phi
\cdot \partial \partial _{\rho }\phi \partial _{\sigma }\partial _{\mu
}\partial _{\nu }\phi ,  \notag \\
&&
\end{eqnarray}%
where $g=\det \left( g_{\mu \nu }\right) $. From these building blocks we
get five independent Lagrangian terms, namely\footnote{%
Note that $\sqrt{\left\vert g\right\vert }\mathcal{I}_{4}^{\left( 3\right) }$
is total derivative so that this tem is absent in $S_{4}$. The factors $%
\left( -1\right) ^{D-1}/2$ are inserted for further convenience.}%
\begin{eqnarray}
S_{4} &=&\int_{M_{\mathbf{\ \mathbb{R}}}^{D}}\sqrt{\mathrm{d}^{D}Z\mathrm{d}%
^{D}\overline{Z}}\sum_{i=1}^{2}B_{4}^{\left( i\right) }\mathcal{I}%
_{4}^{\left( i\right) }  \notag \\
&&+\frac{\left( -1\right) ^{D-1}}{2}\int_{M_{\mathbf{\ \mathbb{R}}}^{D}}%
\mathrm{d}^{D}Z\sum_{i=1}^{3}C_{4}^{\left( i\right) }\mathcal{I}_{4}^{\left(
i\right) }+\frac{\left( -1\right) ^{D-1}}{2}\int_{M_{\mathbf{\ \mathbb{R}}%
}^{D}}\mathrm{d}^{D}\overline{Z}\sum_{i=1}^{3}C_{4}^{\left( i\right) \ast }%
\mathcal{I}_{4}^{\left( i\right) }.
\end{eqnarray}%
Here the coupling constants $B_{4}^{\left( i\right) }$ are real while $%
C_{4}^{\left( i\right) }$ are generally complex. These couplings are not
renormalized for $D>2$ (cf. (\ref{power_counting})). The first two terms are
invariant under the transformation (\ref{hidden symmetry}), while the
remaining two are only form-invariant. The couplings $C_{4}^{\left( i\right)
}$ transform under (\ref{hidden symmetry}) as%
\begin{equation}
C_{4}^{\left( i\right) }\rightarrow C_{4}^{\left( i\right) }\det U\left(
\theta \right) .
\end{equation}%
Using (\ref{inverse_g}), (\ref{I4(1)}), (\ref{I4(2)}) and with help of (\ref%
{dZ_explicit}) and (\ref{det_g}) we obtain for the corresponding Lagrangian $%
\mathcal{L}_{4}$ to the third order in $\phi $ (modulo integration by parts)
\begin{eqnarray}
\mathcal{L}_{4} &=&\frac{1}{\alpha ^{2}}\left( B_{4}^{\left( 1\right)
}+B_{4}^{\left( 2\right) }+\mathrm{Re}C_{4}^{\left( 1\right) }+\mathrm{Re}%
C_{4}^{\left( 2\right) }-\mathrm{Im}C_{4}^{\left( 3\right) }\right) \square
\partial \phi \cdot \square \partial \phi  \label{S_4} \\
&&+\frac{1}{\alpha ^{3}}\left[ \left( \mathrm{Im}C_{4}^{\left( 1\right) }-%
\mathrm{Re}C_{4}^{\left( 3\right) }\right) \square \partial \phi \cdot
\square \partial \phi +\mathrm{Im}C_{4}^{\left( 2\right) }\left\langle
\partial \partial \partial \phi \colon \partial \partial \partial \phi
\right\rangle \right] \square \phi +O\left( \phi ^{4}\right) .  \notag
\end{eqnarray}%
We generate again a higher order kinetic term and an infinite tower of
related interaction terms starting with cubic one. The kinetic term vanishes
for%
\begin{equation}
B_{4}^{\left( 1\right) }+B_{4}^{\left( 2\right) }+\mathrm{Re}C_{4}^{\left(
1\right) }+\mathrm{Re}C_{4}^{\left( 2\right) }-\mathrm{Im}C_{4}^{\left(
3\right) }=0.  \label{kinetic_zero}
\end{equation}%
In such a case, the quartic term\footnote{%
The general quartic interaction term is rather lengthy and is given in the
appendix.} of the Lagrangian $\mathcal{L}_{4}$ reads up to a total derivative%
\begin{eqnarray}
\mathcal{L}_{4}^{\phi ^{4}} &=&\frac{\chi }{2\alpha ^{4}}\left[
6\left\langle \partial \partial \partial \phi \colon \partial \partial
\partial \phi \cdot \partial \partial \phi \cdot \partial \partial \phi
\right\rangle -4\left\langle \square \partial \phi \cdot \partial \partial
\partial \phi \cdot \partial \partial \phi \cdot \partial \partial \phi
\right\rangle \right.  \notag \\
&&\left. -2\square \partial \phi \cdot \partial \partial \phi \cdot \partial
\partial \phi \cdot \square \partial \phi +\left( \square \partial \phi
\cdot \square \partial \phi -\left\langle \partial \partial \partial \phi
\colon \partial \partial \partial \phi \right\rangle \right) \left\langle
\partial \partial \phi \cdot \partial \partial \phi \right\rangle \right]
\notag \\
&&-\frac{\rho }{2\alpha ^{4}}\left( \square \phi \right) ^{2}\square
\partial \phi \cdot \square \partial \phi -\frac{\kappa }{2\alpha ^{4}}%
\left( \square \phi \right) ^{2}\left\langle \partial \partial \partial \phi
\colon \partial \partial \partial \phi \right\rangle ,
\end{eqnarray}%
where $\chi =B_{4}^{\left( 1\right) }+\mathrm{Re}C_{4}^{\left( 1\right) }-%
\mathrm{Im}C_{4}^{\left( 3\right) }$, $\rho =\mathrm{Re}C_{4}^{\left(
1\right) }-\mathrm{Im}C_{4}^{\left( 3\right) }$ and $\kappa =$ $\mathrm{Re}%
C_{4}^{\left( 2\right) }$.

Let us note, that the quartic term $\mathcal{L}_{4}^{\phi ^{4}}$ is very
special. Namely, it is invariant under the quadratic shift $\phi \rightarrow
\phi +\delta _{\theta }^{\mathrm{shift}}\phi $, where
\begin{equation}
\delta _{\theta }^{\mathrm{shift}}\phi \left( x\right) =-\frac{\theta }{2}%
\alpha ^{2}G^{\mu \nu }x_{\mu }x_{\nu },~~~\left( G_{\mu }^{\mu }=0\right) .
\label{quadratic_shift}
\end{equation}%
The reason is as follows. Note that $\delta _{\theta }^{\mathrm{shift}}\phi
\left( x\right) $ is a truncated version of the infinitesimal form (\ref%
{infinitesimal1}) of the symmetry (\ref{hidden symmetry}) of the action $%
S_{4}$ , namely%
\begin{equation}
\delta _{\theta }\phi \left( x\right) =-\frac{\theta }{2}G^{\mu \nu }\left(
\alpha ^{2}x_{\mu }x_{\nu }+\partial _{\mu }\phi \left( x\right) \partial
_{\nu }\phi \left( x\right) \right) ,~~~\left( G_{\mu }^{\mu }=0\right) .
\end{equation}%
Such a transformation converts a general Lagrangian term with $n_{\phi }$
fields into terms with $n_{\phi }-1$ and $n_{\phi }+1$ fields respectively.
Symmetry of the complete action $S_{4}$ under $\phi \rightarrow \phi +$ $%
\delta _{\theta }\phi $ therefore requires cancelations between such terms.
For the the first two terms in the action with lowest $n_{\phi }=n_{\phi
}^{\min },$ $n_{\phi }^{\min }+1$ there are no terms available for the
cancelation of the $n_{\phi }-1$ part of their transforms.\ Because the
latter appear as a result of the truncated transformation $\phi \rightarrow
\phi +\delta _{\theta }^{\mathrm{shift}}\phi $, the first two terms with
lowest $n_{\phi }$ have to be invariant also with respect to the quadtratic
shift symmetry with traceles tensor parameter $G^{\mu \nu }$. For the
general action (\ref{S_4}) this applies to the quadratic and cubic terms,
for the case when (\ref{kinetic_zero}) holds and the quadratic term is
absent, also the quartic term $\mathcal{L}_{4}^{\phi ^{4}}$is invariant
under quadratic shift $\delta _{\theta }^{\mathrm{shift}}\phi $.
\begin{figure}[t]
\begin{center}
\epsfig{width=0.9\textwidth,figure=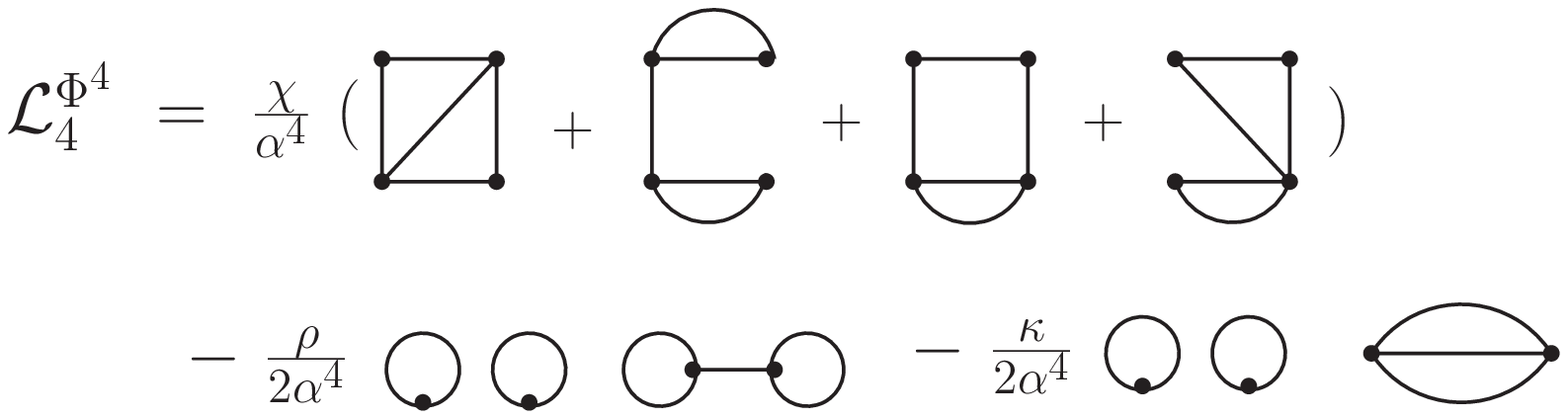}
\end{center}
\caption{The Lagrangian ${\mathcal{L}}_{4}^{\protect\phi ^{4}}$ (in the case
when the higher order kinetic term vanishes) after integration by parts.}
\label{figure_L_4}
\end{figure}

Using integration by parts it can be easily shown that $\mathcal{L}%
_{4}^{\phi ^{4}}$ can be rewritten as%
\begin{eqnarray}
\mathcal{L}_{4}^{\phi ^{4}} &=&\frac{\chi }{\alpha ^{4}}\left(
L_{1}+L_{2}+L_{3}+L_{4}\right)  \notag \\
&&-\frac{\rho }{2\alpha ^{4}}\left( \square \phi \right) ^{2}\square
\partial \phi \cdot \square \partial \phi -\frac{\kappa }{2\alpha ^{4}}%
\left( \square \phi \right) ^{2}\left\langle \partial \partial \partial \phi
\colon \partial \partial \partial \phi \right\rangle ,
\end{eqnarray}%
where $L_{i}$ have been introduced in \cite{Griffin:2014bta} and are
depicted in the figure \ref{figure_L_4} . In such a form $\mathcal{L}%
_{4}^{\phi ^{4}}$fits into the classification of the polynomial shift
symmetric Lagrangians performed in \cite{Hinterbichler:2014cwa} and \cite%
{Griffin:2014bta}. \ The first term corresponds to the Lagrangian denoted in
\cite{Griffin:2014bta} as $(P,N,\Delta )=$ $\left( 2,4,5\right) $, where $P$
is the order of the polynomial shift symmetry, $N=n_{\phi }$ and $\Delta
=n_{\partial }/2$ in terms of our notation. It is invariant under $\delta
_{\theta }^{\mathrm{shift}}\phi $ even for $G_{\mu }^{\mu }\neq 0$. This
behavior can be easily understood within our construction. Note that such a
term is present even for $C_{4}^{\left( i\right) }=0$, when the condition (%
\ref{kinetic_zero}) means $B_{4}^{\left( 2\right) }=-B_{4}^{\left( 1\right)
} $. In such a case, the action $S_{4}$ reduces as a consequence of the
Gauss formula (\ref{gauss_1}) to the Einstein-Hilbert action for the induced
metric $g_{\mu \nu }$
\begin{equation}
S_{4}^{R}=B_{4}^{\left( 1\right) }\sqrt{\left\vert \det \left( g_{\mu \nu
}\right) \right\vert }R.
\end{equation}%
As discussed above, it is invariant under (\ref{hidden symmetry}) without
any restriction on the symmetric tensor $G^{\mu \nu }$.

The remaining two terms of the Lagrangian $\mathcal{L}_{4}^{\phi ^{4}}$ are
both of the general form $\left( \square \phi \right) ^{n}\left( \partial
^{3}\phi \right) ^{2m}$. Such terms have been constructed in \cite%
{Hinterbichler:2014cwa} as the basic interaction terms invariant with
respect to quadratic shift $\delta _{\theta }^{\mathrm{shift}}\phi $ with
traceless $G^{\mu \nu }$.

\subsection{Lagrangians with $\protect\delta \geq 6$ and a role of Lovelock
invariants}

The first Lagrangian $\mathcal{L}_{\delta }$ which is renormalized by loop
corrections in $\ D=4$ dimensions (and which is not renormalized for $D>4$)
corresponds to $\delta =6$, i.e. $n_{\partial }=2n_{\phi }+4$. According to (%
\ref{index}) the invariants $\mathcal{L}_{Z}$, $\mathcal{L}_{\overline{Z}}$
and $\mathcal{L}_{Z\overline{Z}}$ are linear combinations of the terms with $%
n_{\mathcal{K}}+n_{D}=4$, schematically
\begin{equation}
\mathcal{K}^{4},~~D\mathcal{K}^{3},~~D^2\mathcal{K}^2,~~D^3\mathcal{K}.
\end{equation}%
There is a large (but finite) number of combinations how to contract the
indices of the above structures with the inverse metric $g^{\mu \nu }$; we
will not list them explicitly.

As discussed above, for each resulting independent Lagrangian, the first two
terms with with lowest $n_{\phi }=n_{\phi }^{\min },$ $n_{\phi }^{\min }+1~$%
have to be invariant with respect to the quadratic shift (\ref%
{quadratic_shift}). Such terms (provided they are invariant also for $G_{\mu
}^{\mu }\neq 0$, i.e. if they originate in $\mathcal{L}_{Z\overline{Z}}$)
correspond to Lagrangians $(P,N,\Delta )=(2,n_{\phi },n_{\phi }+2)$ (using
the nomenclature of \cite{Griffin:2014bta}). Here the most interesting are
those with maximal posible $n_{\phi }^{\min }$, i.e. those with minimal
number of derivatives per field (especially $n_{\partial }<3n_{\phi }$).
Naively $n_{\phi }^{\min }=n_{\mathcal{K}}$ and therefore we apparently
cannot go beyond the uninteresting $(P,N,\Delta )=(2,4,6)$ terms with three
derivatives per field, which originate from the $\mathcal{K}^{4}=O\left(
\phi ^{4}\right) $ part of the general action. However, in analogy with the $%
\delta =4$ case, for $D>4$ there exist a special combination of the $%
\mathcal{K}^{4}$ invariants for which $n_{\phi }^{\min }=6$. Not
accidentally such a combination of $\mathcal{K}^{4}$ terms is the Lovelock
invariant \cite{Lovelock:1971yv}%
\begin{eqnarray}
\mathcal{L}_{Z\overline{Z}}^{R^{2}} &=&R^{2}-4R_{\alpha \beta }R^{\alpha
\beta }+R_{\alpha \beta \mu \nu }R^{\alpha \beta \mu \nu }  \notag \\
&=&g^{\rho \sigma }g^{\kappa \lambda }\left( \mathcal{K}_{\rho \mu \alpha }%
\mathcal{K}_{\sigma \nu \beta }-\mathcal{K}_{\rho \mu \beta }\mathcal{K}%
_{\sigma \nu \alpha }\right) \left( \mathcal{K}_{\kappa \omega \gamma }%
\mathcal{K}_{\lambda \tau \delta }-\mathcal{K}_{\kappa \omega \delta }%
\mathcal{K}_{\lambda \tau \gamma }\right)  \notag \\
&&\times \left[ g^{\alpha \mu }g^{\gamma \omega }g^{\beta \nu }g^{\tau
\delta }-4g^{\alpha \mu }g^{\gamma \omega }g^{\beta \delta }g^{\nu \tau
}+g^{\alpha \gamma }g^{\beta \delta }g^{\mu \omega }g^{\nu \tau }\right] .
\end{eqnarray}%
The $n_{\phi }=6$ term of the invariant action
\begin{equation}
S_{6}^{R^{2}}\equiv B_{6}^{R^{2}}\int_{M_{\mathbf{\ \mathbb{R}}}^{D}}\sqrt{%
\mathrm{d}^{D}Z\mathrm{d}^{D}\overline{Z}}\mathcal{L}_{Z\overline{Z}}^{R^{2}}
\end{equation}%
built from the Lovelock invariant corresponds then to quadratic shift
symmetric term $(P,N,\Delta )=(2,6,8)$.

For $\delta =2n+2>6$ we meet an analogous situation, namely the terms $%
\mathcal{K}^{2n}$ can be combined for $D>\delta -2=2n$ into the Lovelock
invariant $\mathcal{L}_{Z\overline{Z}}^{R^{n}}$ where
\begin{equation}
\mathcal{L}_{Z\overline{Z}}^{R^{n}} =\frac{1}{2^{n}}\delta _{\alpha
_{1}\beta _{1}}^{\mu _{1}\nu _{1}}\ldots _{\alpha _{n}\beta _{n}}^{\mu
_{n}\nu _{n}}R_{\mu _{1}\nu _{1}}^{~~~~\alpha _{1}\beta _{1}}\ldots R_{\mu
_{n}\nu _{n}}^{~~~~\alpha _{n}\beta _{n}} ,
\end{equation}
where the generalized Kronecker delta is
\begin{equation}
\delta _{\alpha _{1}}^{\mu _{1}}\ldots _{\alpha _{n}}^{\mu _{n}}=\frac{1}{n!}%
\delta _{\lbrack \alpha _{1}}^{\mu _{1}}\delta _{\alpha _{2}}^{\mu
_{2}}\ldots \delta _{\alpha _{n}]}^{\mu _{n}}.
\end{equation}%
Then the corresponding invariant action
\begin{equation}
S_{\delta =2n+2}^{R^{n}}=B_{\delta }^{R^{n}}\int_{M_{\mathbf{\ \mathbb{R} }%
}^{D}}\sqrt{\mathrm{d}^{D}Z\mathrm{d}^{D}\overline{Z}}\mathcal{L}_{Z%
\overline{Z}}^{R^{n}}
\end{equation}%
starts with a term with $n_{\phi }=\delta =2n+2$. \ Indeed, using\footnote{%
On the right hand side the indices are rised by the flat metric $\eta ^{\mu
\nu }$.}
\begin{equation}
R_{\mu \nu }^{~~~\alpha \beta }=\frac{1}{\alpha ^{2}}\left( \partial ^{\beta
}\partial _{\nu }\partial \phi \cdot \partial \partial _{\mu }\partial
^{\alpha }\phi -\partial ^{\alpha }\partial _{\nu }\partial \phi \cdot
\partial \partial _{\mu }\partial ^{\beta }\phi \right) +O\left( \phi
^{4}\right)
\end{equation}%
and expressing $S_{\delta =2n+2}^{R^{n}}$ in terms of the Galileon field we
get%
\begin{eqnarray}
&&\sqrt{\left\vert \det \left( g_{\mu \nu }\right) \right\vert }\mathcal{L}%
_{Z\overline{Z}}^{R^{n}}=\frac{1}{\alpha ^{2n}}\delta _{\alpha _{1}\beta
_{1}}^{\mu _{1}\nu _{1}}\ldots _{\alpha _{n}\beta _{n}}^{\mu _{n}\nu
_{n}}\prod\limits_{i=1}^{n}\partial ^{\beta _{i}}\partial _{\nu
_{i}}\partial \phi \cdot \partial \partial _{\mu _{i}}\partial ^{\alpha
_{\iota }}\phi +O\left( \phi ^{2n+2}\right)  \notag \\
&=&\frac{1}{\alpha ^{2n}}\delta _{\alpha _{1}\beta _{1}}^{\mu _{1}\nu
_{1}}\ldots _{\alpha _{n}\beta _{n}}^{\mu _{n}\nu _{n}}\partial _{\mu
_{1}}\partial ^{\alpha _{1}}\left[ \partial ^{\beta _{1}}\partial _{\nu
_{1}}\partial \phi \cdot \partial \phi \prod\limits_{i=2}^{n}\partial
^{\beta _{i}}\partial _{\nu _{i}}\partial \phi \cdot \partial \partial _{\mu
_{i}}\partial ^{\alpha _{\iota }}\phi \right] +O\left( \phi ^{2n+2}\right)
\notag \\
&&
\end{eqnarray}%
where we have used the antisymmetry of $\delta _{\alpha _{1}\beta _{1}}^{\mu
_{1}\nu _{1}}\ldots _{\alpha _{n}\beta _{n}}^{\mu _{n}\nu _{n}}$. The first
term is a total derivative and thus $S_{\delta =2n+2}^{R^{n}}=O\left( \phi
^{2n+2}\right) $

For general $\delta =2n+2,$ for which $n_{\partial }=\delta +2n_{\phi
}-2=2n+2n_{\phi }$ and for $D>\delta -2=2n$ we can therefore generate the
quadratic shift invariants $(P,N,\Delta )=(2,2n+2,3n+2)$ as the sum of the
lowest $n_{\phi }=2n+2$ terms of the Lovelock action $S_{\delta
=2n+2}^{R^{n}}$. Of course, the complete Lovelock action is symmetric with
respect to the complete transformation (\ref{hidden symmetry}).

\section{Note on the analytic continuation $\protect\alpha \rightarrow
\mathrm{i}a$\label{section_analytic_continuation}}

Let comment briefly on the case of imaginary $\alpha $, i.e. when the coset
space transformation is formulated in terms of the real coordinates $Z^{\pm
\mu }$ (see (\ref{duality_imaginary})). In such a case we can repeat the
above brane constructions almost literaly, however the target space in not a
K\"{a}hler manifold any more. Instead, we start with a flat space with the
metric with signature $\left( 2D,2D\right) $%
\begin{equation}
\mathrm{d}s^{2}=\mathrm{d}Z^{+}\cdot \mathrm{d}Z^{-}=\eta _{\mu \nu }\left(
\mathrm{d}X^{\mu }\mathrm{d}X^{\nu }-\frac{1}{a^{2}}\mathrm{d}L^{\mu }%
\mathrm{d}L^{\nu }\right)
\end{equation}%
and the form $\omega $ we replace with%
\begin{equation}
\omega =-\frac{1}{2}\eta _{\mu \nu }\mathrm{d}Z^{+\mu }\wedge \mathrm{d}%
Z^{-\nu }=\frac{1}{a}\eta _{\mu \nu }\mathrm{d}X^{\mu }\wedge \mathrm{d}%
L^{\nu }.
\end{equation}%
The group of simultaneous symmetry of both these structures is now $%
%TCIMACRO{\U{211d} }%
%BeginExpansion
\mathbb{R}
%EndExpansion
^{2D}\rtimes O(D,D)\cap Sp\left( 2D\right) $. The induced metric on the
brane reads in this case%
\begin{equation}
g_{\mu \nu }=\eta _{\mu \nu }-\frac{1}{a^{2}}\partial _{\mu }\partial \phi
\cdot \partial \partial _{\nu }\phi .  \label{induced_metric_real}
\end{equation}%
and the other invariant building blocks derived from $g_{\mu \nu }$ can be
easily obtained from the previous case by means of analytic continuation $%
\alpha \rightarrow \mathrm{i}a$; e.g. the Christoffel symbols of the second
kind are now%
\begin{equation}
\Gamma _{\rho \sigma \mu }=-\frac{1}{a^{2}}\partial _{\mu }\partial _{\sigma
}\partial \phi \cdot \partial \partial _{\rho }\phi .
\end{equation}%
The normal vectors $n_{a}^{A}$ are again constructed in terms of $D-$bein $%
m_{a}$ with respect to the induced metric (\ref{induced_metric_real}), (cf.
also (\ref{m_a_normalization})) as
\begin{equation}
n_{a}^{A}=\left( \frac{1}{a}\partial ^{\alpha }\partial \phi \cdot
m_{a},am_{a}^{\beta }\right) ,
\end{equation}%
and are normalized according to\footnote{%
The minus sign on the right hand side of the first normalization condition
reflects the different signature of the target space metric $G_{AB}$ in
comparisson with the real $\alpha $ case.}%
\begin{equation}
G_{AB}n_{a}^{A}n_{b}^{B}=-\eta _{ab},~~~G_{AB}n_{a}^{A}e_{\beta }^{B}=0.
\end{equation}%
The dual basis reads then%
\begin{equation}
e_{A}^{\mu }=g^{\mu \sigma }\left( \eta _{\sigma \alpha },-\frac{1}{a^{2}}%
\partial _{\sigma }\partial _{\beta }\phi \right) ,~~~n_{A}^{a}=\frac{1}{a}%
\eta ^{ab}\left( m_{b}\cdot \partial \partial _{\alpha }\phi ,\eta _{\beta
\sigma }m_{b}^{\sigma }\right) ,
\end{equation}%
and thus the extrinsic curvature $K_{\mu \nu }^{a}$ and twist connection are
now%
\begin{eqnarray}
K_{\mu \nu }^{a} &=&-\frac{1}{a}\eta ^{ab}m_{b}\cdot \partial \partial _{\mu
}\partial _{\nu }\phi \\
\beta _{\mu a}^{b} &=&\eta ^{bc}\left( m_{c}^{\alpha }g_{\alpha \beta
}\partial _{\mu }m_{a}^{\beta }-\frac{1}{a^{2}}m_{c}\cdot \partial \partial
\phi \cdot \partial \partial _{\mu }\partial \phi \cdot m_{\alpha }\right) .
\end{eqnarray}%
Again, as a consequence of the Gauss-Codazzi relations the basic building
blocks for the higher order lagrangians can be taken to be the induce metric
$g_{\mu \nu }$, its inverse $g^{\mu \nu }$ and the covariant derivatives of
the tensor

\begin{equation*}
\mathcal{K}_{\alpha \mu \nu }=g_{\alpha \beta }m_{a}^{\beta }K_{\mu \nu
}^{a}=-\frac{1}{a}\partial _{\alpha }\partial _{\mu }\partial _{\nu }\phi
\end{equation*}%
together with measures $\mathrm{d}^{D}Z^{+}$, $\mathrm{d}^{D}Z^{-}$ and \ $%
\mathrm{d}^{D}x\sqrt{\left\vert \det \left( g_{\mu \nu }\right) \right\vert }%
\equiv \sqrt{\mathrm{d}^{D}Z^{+}\mathrm{d}^{D}Z^{-}}$. The invariant action
has the general form
\begin{equation}
S_{inv}=\int_{M_{\mathbf{\ \mathbb{R}}}^{D}}\mathrm{d}^{D}Z\mathcal{L}%
_{Z^{+}}+\int_{M_{\mathbf{\ \mathbb{R}}}^{D}}\mathrm{d}^{D}Z^{-}\mathcal{L}%
_{Z^{-}}+\int_{M_{\mathbf{\ \mathbb{R}}}^{D}}\sqrt{\mathrm{d}^{D}Z^{+}%
\mathrm{d}^{D}Z^{-}}\mathcal{L}_{Z^{+}Z^{-}}.
\end{equation}%
where in contrast to the case of real $\alpha $ the invariants $\mathcal{L}%
_{Z^{+}}$ and $\mathcal{L}_{Z^{-}}$ are not correlated.

\section{Conclusion}

The aim of this work was to discuss in detail the symmetry properties of the
Special Galileon in flat space. We have found that the hidden symmetry which
dictates the form of its Lagrangian and which is responsible for its
peculiar $O\left( p^{3}\right) $ soft limit behavior can be understood as a
special case of more general family of duality transformations. Similarly,
the Special Galileon itself turns out to be a special member of a wider
family of theories which are interrelated by the duality transformations
mentioned above. These originate in special transformation of the coset
space $GAL\left( D,1\right) /SO\left( 1,D-1\right) $. There are two branches
of such transformations which can be parameterized either by the matrix $%
U=\exp \left( \mathrm{i}\mathcal{G}\right) $ or by the matrix $U=\exp \left(
\mathcal{G}\right) $ where the real rank two tensor $\mathcal{G}_{~\nu
}^{\mu }$ satisfies $\eta _{\mu \rho }\mathcal{G}_{~\nu }^{\rho }=\eta _{\nu
\rho }\mathcal{G}_{~\mu }^{\rho }$. The hidden Galileon symmetry found
previously in \cite{Hinterbichler:2015pqa} is then identified with an
infinitesimal form of this symmetries for $\mathrm{Tr}\mathcal{G}=0$ after
imposing the Inverse Higgs Constraint.

As we have shown, the Special Galileon in $D$ dimensions can be also
interpreted within a different geometrical language. Within such a
framework, the Galileon field $\phi $ is a scalar degree of freedom
describing (in particular gauge) the position of the $D-$dimensional probe
brane embedded in a $2D-$dimensional target space. The Inverse Higgs
Constraint can be formulated geometrically as a constraint on the brane
embedding, namely forcing an appropriate target space two-form $\omega $ to
vanish when restricted to the brane. The ordinary Galileon symmetry and the
symmetry/duality of Special Galileon can be then identified with a subgroup
of isometries of the target space which are at the same time symmetries of
the form $\omega $. For the two branches of the Special Galileon, the metric
of the target space is flat with signature either $(2,2D-2)$ or $\left(
2D,2D\right) $. In the first case, the target space has a natural structure
of a K\"{a}hler manifold and the group of Special Galileon symmetry/duality
consists of symmetries of the corresponding hermitian form, namely $\mathbb{R%
} ^{2D}\rtimes O\left( 2,D-2\right) \cap Sp\left( 2D\right) \approx \mathbb{C%
}^{D}\rtimes U(1,D-1)$. The complex translations represent spatial
translation and the Galileon linear shift symmetries, while the $U(1,D-1)$
transformations can be interpreted as Lorentz transformations and hidden
symmetry/duality of the Special Galileon. In the second case, the group is $%
\mathbb{R} ^{2D}\rtimes O(D,D)\cap Sp\left( 2D\right) $ and the target space
has not any compatible complex structure.

The Lagrangians which are symmetric/dual with respect to these symmetries
can be then constructed according the general probe brane prescription. The
building blocks can be reduced, as a consequence of the Gauss-Codazzi
formulae, to the induced metric on the brane, the extrinsic curvature tensor
and its covariant derivatives as well as invariant/covariant measures on the
brane. This allows to reconstruct the Special Galileon action, and more
importantly simplifies considerably the classification of the higher order
counterterms. We have performed such classification up to the invariants of
the schematic form $\partial ^{2n+2}\phi ^{n}$ ($n$ arbitrary).

As a byproduct we have established a close relations between such higher
order invariants and the Lagrangians invariant with respect to the quadratic
shift symmetry. We have found, that for each higher order invariant, which
in general consist of infinite tower of terms, the sum of vertices with
minimal (and next to minimal) number of fields is automatically quadratic
shift invariant. Moreover, the invariants constructed as the Lovelock terms,
schematically $R^{n}$ in $D>2n$ dimensions, where $R$ stays for intrinsic
curvature tensor on the brane, allow us to easily obtain the $(P,N,\Delta
)=(2,2n+2,3n+2)$ polynomial shift invariant Lagrangians.

\bigskip

\bigskip \textbf{Acknowledgment} The author would like to thank to Karol
Kampf for discussions. This work is supported in part by Czech Government
project no. LH14035 and GA\v{C}R 15-18080S.

\bigskip \appendix

\section{Invariance of IHC under special Galileon duality\label{IHC1}}

Here we give a proof of the invariance of the IHC constraint under the
transformation (\ref{hidden symmetry}). Let us calculate the transformed
form $\left[ \omega _{A}\right] _{\theta }$ defined as%
\begin{equation*}
\left[ \omega _{A}\right] _{\theta }=\mathrm{d}\phi _{\theta }+\mathrm{i}%
\frac{\alpha }{4}\left( Z_{\theta }-\overline{Z}_{\theta }\right) ^{T}\cdot
\eta \cdot \left( \mathrm{d}Z_{\theta }+\mathrm{d}\overline{Z}_{\theta
}\right)
\end{equation*}%
where
\begin{eqnarray}
Z_{\theta } &=&U\left( \theta \right) \cdot Z,~~~~~\overline{Z}_{\theta
}=U\left( -\theta \right) \cdot \overline{Z}  \notag \\
\phi _{\theta } &=&\phi +\mathrm{i}\frac{\alpha }{8}\left( Z^{2}-\overline{Z}%
^{2}\right) -\mathrm{i}\frac{\alpha }{8}\left( Z_{\theta }^{2}-\overline{Z}%
_{\theta }^{2}\right)
\end{eqnarray}%
Inserting this into $\left[ \omega _{A}\right] _{\theta }$ we get
\begin{eqnarray*}
\left[ \omega _{A}\right] _{\theta } &=&\mathrm{d}\phi +\mathrm{i}\frac{%
\alpha }{4}\left( Z^{T}\cdot \eta \cdot \mathrm{d}Z-\overline{Z}^{T}\cdot
\eta \cdot \mathrm{d}\overline{Z}\right) \\
&&-\mathrm{i}\frac{\alpha }{4}\left( Z_{\theta }^{T}\cdot \eta \cdot \mathrm{%
d}Z_{\theta }-\overline{Z}_{\theta }^{T}\cdot \eta \cdot \mathrm{d}\overline{%
Z}_{\theta }\right) \\
&&+\mathrm{i}\frac{\alpha }{4}\left( Z_{\theta }-\overline{Z}_{\theta
}\right) ^{T}\cdot \eta \cdot \left( \mathrm{d}Z_{\theta }+\mathrm{d}%
\overline{Z}_{\theta }\right) \\
&=&\mathrm{d}\phi +\mathrm{i}\frac{\alpha }{4}\left( Z^{T}\cdot \eta \cdot
\mathrm{d}Z-\overline{Z}^{T}\cdot \eta \cdot \mathrm{d}\overline{Z}\right) \\
&&+\mathrm{i}\frac{\alpha }{4}\left( Z_{\theta }^{T}\cdot \eta \cdot \mathrm{%
d}\overline{Z}_{\theta }-\overline{Z}_{\theta }^{T}\cdot \eta \cdot \mathrm{d%
}Z_{\theta }\right)
\end{eqnarray*}%
Because%
\begin{equation}
\mathrm{d}Z_{\theta }=U\left( \theta \right) \cdot \mathrm{d}Z,~~~~\mathrm{d}%
\overline{Z}_{\theta }=U\left( -\theta \right) \cdot \mathrm{d}\overline{Z}
\notag
\end{equation}%
and%
\begin{equation*}
U\left( \theta \right) ^{T}\cdot \eta \cdot U\left( -\theta \right) =U\left(
-\theta \right) ^{T}\eta \cdot U\left( \theta \right) =\eta
\end{equation*}%
we get%
\begin{equation*}
Z_{\theta }^{T}\cdot \eta \cdot \mathrm{d}\overline{Z}_{\theta }=Z^{T}\cdot
\eta \cdot \mathrm{d}\overline{Z},~~~,\overline{Z}_{\theta }^{T}\cdot \eta
\cdot \mathrm{d}Z_{\theta }=\overline{Z}^{T}\cdot \eta \cdot \mathrm{d}Z
\end{equation*}%
and therefore%
\begin{equation*}
\left[ \omega _{A}\right] _{\theta }=\mathrm{d}\phi _{\theta }+\mathrm{i}%
\frac{\alpha }{4}\left( Z-\overline{Z}\right) ^{T}\cdot \eta \cdot \left(
\mathrm{d}Z+\mathrm{d}\overline{Z}\right) =\omega _{A}.
\end{equation*}

\section{Invariance of IHC under generalized special Galileon duality\label%
{IHC2}}

In this appendix we demonstrate invariance of the IHC constraint under (\ref%
{generalized_special_duality}). For this purpose we rewite (\ref%
{generalized_special_duality}) in the form%
\begin{eqnarray}
Z_{\theta }^{+} &=&Z^{+}=x+\frac{1}{a}L  \label{krok1} \\
L-L_{\theta } &=&\theta G\left( Z^{+}\right) \cdot Z^{+}  \label{krok2} \\
\phi _{\theta } &=&\phi +\frac{1}{2a}L^{2}-\frac{1}{2a}L_{\theta }^{2}-\frac{%
\theta }{N}Z^{+}\cdot G\left( Z^{+}\right) \cdot Z^{+}  \label{krok3}
\end{eqnarray}%
where we abreviated%
\begin{equation*}
G\left( Z^{+}\right) ^{\mu \nu }=G^{\mu \nu \alpha _{1}\alpha _{2}\ldots
\alpha _{N-2}}Z_{\alpha _{1}}^{+}Z_{\alpha _{2}}^{+}\ldots Z_{\alpha
_{N-2}}^{+}.
\end{equation*}%
Inserting (\ref{krok3}) into
\begin{equation*}
\left[ \omega _{A}\right] _{\theta }=\mathrm{d}\phi _{\theta }-L_{\theta
}\cdot \mathrm{d}x_{\theta }
\end{equation*}%
we get%
\begin{eqnarray*}
\left[ \omega _{A}\right] _{\theta } &=&\mathrm{d}\phi +\frac{1}{a}L\cdot
\mathrm{d}L-\frac{1}{a}L_{\theta }\cdot \mathrm{d}L_{\theta }-\theta
Z^{+}\cdot G\left( Z^{+}\right) \cdot \mathrm{d}Z^{+}-L_{\theta }\cdot
\mathrm{d}x_{\theta } \\
&=&\mathrm{d}\phi +\frac{1}{a}L\cdot \mathrm{d}L-L_{\theta }\cdot \mathrm{d}%
Z_{\theta }^{+}-\left( L-L_{\theta }\right) \cdot \mathrm{d}Z^{+} \\
&=&\mathrm{d}\phi +\frac{1}{a}L\cdot \mathrm{d}L-L_{\theta }\cdot \mathrm{d}%
Z^{+}-\left( L-L_{\theta }\right) \cdot \mathrm{d}Z^{+} \\
&=&\mathrm{d}\phi +\frac{1}{a}\left( L-L_{\theta }\right) \cdot \mathrm{d}%
L-L_{\theta }\cdot \mathrm{d}x-\left( L-L_{\theta }\right) \cdot \mathrm{d}%
Z^{+} \\
&=&\mathrm{d}\phi -\left( L-L_{\theta }\right) \cdot \mathrm{d}x-L_{\theta
}\cdot \mathrm{d}x=\mathrm{d}\phi -L\cdot \mathrm{d}x=\omega _{A}
\end{eqnarray*}%
where we have used (\ref{krok1}) and (\ref{krok2}) to get the second line, (%
\ref{krok1}) to get the third line and then twice (\ref{krok1}).

\section{Explicit form of the quartic term of $\protect\delta =4$ Lagrangian
\label{quartic_lagrangian}}

In this appendix we give explicit form of the quartic term of the most
general $\delta =4$ Lagrangian
\begin{eqnarray}
\mathcal{L}_{4}^{\phi ^{4}} &=&\frac{\chi }{2\alpha ^{4}}\left[
6\left\langle \partial \partial \partial \phi \colon \partial \partial
\partial \phi \cdot \partial \partial \phi \cdot \partial \partial \phi
\right\rangle -4\left\langle \square \partial \phi \cdot \partial \partial
\partial \phi \cdot \partial \partial \phi \cdot \partial \partial \phi
\right\rangle \right.  \notag \\
&&\left. -2\square \partial \phi \cdot \partial \partial \phi \cdot \partial
\partial \phi \cdot \square \partial \phi +\left( \square \partial \phi
\cdot \square \partial \phi -\left\langle \partial \partial \partial \phi
\colon \partial \partial \partial \phi \right\rangle \right) \left\langle
\partial \partial \phi \cdot \partial \partial \phi \right\rangle \right]
\notag \\
&&-\frac{\rho }{2\alpha ^{4}}\left( \square \phi \right) ^{2}\square
\partial \phi \cdot \square \partial \phi -\frac{\kappa }{2\alpha ^{4}}%
\left( \square \phi \right) ^{2}\left\langle \partial \partial \partial \phi
\colon \partial \partial \partial \phi \right\rangle  \notag \\
&&+\frac{\xi }{2\alpha ^{4}}\left[ \left\langle \partial \partial \partial
\phi \colon \partial \partial \partial \phi \right\rangle \left\langle
\partial \partial \phi \cdot \partial \partial \phi \right\rangle
-6\left\langle \partial \partial \partial \phi \colon \partial \partial
\partial \phi \cdot \partial \partial \phi \cdot \partial \partial \phi
\right\rangle \right] ,
\end{eqnarray}%
where we abreviated%
\begin{eqnarray}
\xi &=&B_{4}^{\left( 1\right) }+B_{4}^{\left( 2\right) }+\mathrm{Re}%
C_{4}^{\left( 1\right) }+\mathrm{Re}C_{4}^{\left( 2\right) }-\mathrm{Im}%
C_{4}^{\left( 3\right) }  \notag \\
\chi &=&B_{4}^{\left( 1\right) }+\mathrm{Re}C_{4}^{\left( 1\right) }-\mathrm{%
Im}C_{4}^{\left( 3\right) },  \notag \\
\rho &=&\mathrm{Re}C_{4}^{\left( 1\right) }-\mathrm{Im}C_{4}^{\left(
3\right) },~~~\kappa =\mathrm{Re}C_{4}^{\left( 2\right) }
\end{eqnarray}

\section{Transformation of the actions with respect to $GL\left( 2, \mathbb{R%
} \right) $ duality\label{duality_on_special_galileon}}

As we have discussed in the main text, not all the Galileon actions have
nontrivial physical content. As we have mentioned, some of them might be
transformed into free theory or to the tadpole term by means of the $%
GL\left( 2, \mathbb{R} \right) $duality (\ref{old_duality}) mentioned in
section \ref{coset and dualities}. In this appendix we give a proof of such
relations in more detail.

The Lagrangians under consideration can be written as%
\begin{equation}
\mathcal{L}=\mathcal{N}\sum_{n}d_{n}\mathcal{L}_{n}
\end{equation}%
where $\mathcal{N}$ is a real overall constant (which may be needed e.g. in
order to get a canonical normalization of the kinetic term). Under the
duality transformation (\ref{old_duality})
\begin{equation}
x_{\theta }=x-2\theta \partial \phi \left( x\right) ,~~~~\phi _{\theta
}\left( x_{\theta }\right) =\phi \left( x\right) -\theta \partial \phi
\left( x\right) \cdot \partial \phi \left( x\right)
\end{equation}%
the constants $d_{n}$ change according to the relation (see \cite%
{Kampf:2014rka} for details)%
\begin{equation}
d_{n}\left( \theta \right) =\frac{1}{n}\sum_{m=1}^{n}m\left(
\begin{array}{c}
D-m+1 \\
n-m%
\end{array}%
\right) \left( -2\theta \right) ^{n-m}d_{m}  \label{d_space_duality}
\end{equation}%
Paricularly we get%
\begin{equation}
d_{1}\left( \theta \right) =d_{1}.
\end{equation}%
For the case of the action $S(\alpha ,\beta )$ (see (\ref{action_alpha_xi}%
))we have%
\begin{equation*}
d_{n}\left( \alpha ,\beta \right) =\frac{1}{2\mathrm{i}n}\alpha \mathrm{e}^{%
\mathrm{i}\beta }\left(
\begin{array}{c}
D \\
n-1%
\end{array}%
\right) \left( \frac{\mathrm{i}}{\alpha }\right) ^{n-1}+h.c.
\end{equation*}%
Therefore inserting this in \ref{d_space_duality}() for $n>1$ we get%
\begin{eqnarray}
d_{n}\left( \theta \right) &=&\frac{1}{2\mathrm{i}n}\alpha \mathrm{e}^{%
\mathrm{i}\beta }\frac{\left( -2\theta \right) ^{n-1}}{\left( D-n+1\right) !}%
\sum_{m=1}^{n}\frac{D!}{\left( n-m\right) !\left( m-1\right) !}\left( -\frac{%
\mathrm{i}}{2\theta \alpha }\right) ^{m-1}+h.c.  \notag \\
&=&\frac{1}{2\mathrm{i}n}\alpha \mathrm{e}^{\mathrm{i}\beta }\left(
\begin{array}{c}
D \\
n-1%
\end{array}%
\right) \left( -2\theta \right) ^{n-1}\sum_{m=0}^{n-1}\left(
\begin{array}{c}
n-1 \\
m%
\end{array}%
\right) \left( -\frac{\mathrm{i}}{2\theta \alpha }\right) ^{m}+h.c.  \notag
\end{eqnarray}%
As a result we obtain
\begin{eqnarray}
d_{n}\left( \theta \right) &=&\frac{1}{2\mathrm{i}n}\alpha \mathrm{e}^{%
\mathrm{i}\beta }\left(
\begin{array}{c}
D \\
n-1%
\end{array}%
\right) \left( \frac{\mathrm{i}}{\alpha }-2\theta \right) ^{n-1}+h.c. \\
&=&\frac{1}{2\mathrm{i}n}\alpha \mathrm{e}^{\mathrm{i}\left( \beta -\left(
n-1\right) \beta _{\theta }\right) }\left(
\begin{array}{c}
D \\
n-1%
\end{array}%
\right) \left( \frac{\mathrm{i}}{\alpha _{\theta }}\right) ^{n-1}+h.c.,
\end{eqnarray}%
where we expressed%
\begin{equation}
\frac{\mathrm{i}}{\alpha }-2\theta =\frac{\mathrm{i}}{\alpha _{\theta }}%
\mathrm{e}^{\mathrm{i}\beta _{\theta }}
\end{equation}%
in terms of real $\alpha _{\theta }>0$ and $-3\pi /2<\beta _{\theta }<\pi /2$
for finite $\theta \neq 0$. \ For $n$ odd we can force $d_{n}\left( \theta
\right) $ to vanish for
\begin{equation}
\sin \left( \beta -\left( n-1\right) \beta _{\theta }\right) =0
\end{equation}%
For $n$ even we can do the same provided
\begin{equation}
\cos \left( \beta -\left( n-1\right) \beta _{\theta }\right) =0.
\end{equation}

For the action $S_{\pm }\left( a,c_{\pm }\right) $ (see \ (\ref{s_plus_minus}%
))we have%
\begin{equation}
d_{n}^{\pm }\left( a,c_{\pm }\right) =\frac{1}{n}\mathrm{e}^{c_{\pm }}\left(
\begin{array}{c}
D \\
n-1%
\end{array}%
\right) \frac{\left( -1\right) ^{n}}{a^{n-2}}
\end{equation}%
and repating the above calculation with imaginary $\alpha =\pm \mathrm{i}a$
and $\beta =c_{\pm }/\mathrm{i}$ we get finally for $n>1$%
\begin{equation}
d_{n}^{\pm }\left( \theta \right) =\frac{1}{2n}a\mathrm{e}^{c_{\pm }}\left(
\begin{array}{c}
D \\
n-1%
\end{array}%
\right) \left( \pm \frac{1}{a}-2\theta \right) ^{n-1}.
\label{d_n_theta_plus_minus}
\end{equation}%
Therefore for $\theta =\pm 1/2a$ we can made all the coefficients with $n>1$
vanish. The theory described with action $S_{+}(a,c_{+})$ or $S_{-}\left(
a,c_{-}\right) $ is therefore dual to the action
\begin{equation*}
S_{\pm }\left( a,c_{\pm }\right) \overset{\theta =\pm 1/2a}{\rightarrow }%
\frac{\left( -1\right) ^{D-1}D!}{2}a\mathrm{e}^{c_{\pm }}\int \mathrm{d}%
^{D}x\phi
\end{equation*}%
with only the tadpole term.

Finally, let as assume the action $S^{L}$ (see (\ref{S_L})) with
\begin{equation}
d_{1}^{L}\left( a\right) =0,~~~d_{n}^{L}\left( a\right) =\frac{\left(
-1\right) ^{D-1}}{n}\left(
\begin{array}{c}
D-1 \\
n-2%
\end{array}%
\right) \frac{1}{a^{n-2}}.
\end{equation}%
Note that we can write%
\begin{equation*}
d_{n}^{L}\left( a\right) =-\frac{\left( -1\right) ^{D-1}}{D}a^{2}\frac{%
\partial }{\partial a}\left[ \frac{1}{a}\frac{1}{n}\left(
\begin{array}{c}
D \\
n-1%
\end{array}%
\right) \frac{1}{a^{n-2}}\right]
\end{equation*}%
and therefore using (\ref{d_n_theta_plus_minus}) we get for $n=1,2$%
\begin{equation*}
d_{1}^{L}\left( \theta \right) =0,~~~d_{2}^{L}\left( \theta \right) =\frac{%
\left( -1\right) ^{D-1}}{2}
\end{equation*}
while for $n>2$%
\begin{eqnarray*}
d_{n}^{L}\left( \theta \right) &=&-\frac{\left( -1\right) ^{D-1}}{D}a^{2}%
\frac{\partial }{\partial a}\left\{ \frac{1}{a}\left[ \frac{1}{n}a\left(
\begin{array}{c}
D \\
n-1%
\end{array}%
\right) \left( \frac{1}{a}-2\theta \right) ^{n-1}\right] \right\} \\
&=&\frac{\left( -1\right) ^{D-1}}{D}\frac{1}{n}\left(
\begin{array}{c}
D \\
n-1%
\end{array}%
\right) \left( \frac{1}{a}-2\theta \right) ^{n-2}.
\end{eqnarray*}%
The choice $\theta =1/2a$ transforms $S_{L}$ to the free action with only
the kinetic term. Therefore any linear combination of the actions $S_{+}$
and $S^{L}$ can be simultaneously transformed to the free theory with
tadpole term.

\bigskip

\bibliographystyle{utphys}
\bibliography{sgal}
{}

\end{document}